\documentclass[twocolumn,10pt]{IEEEtran}
\usepackage{ifpdf, flushend}
\usepackage{subfigure}
\usepackage{braket}

\ifCLASSINFOpdf
  \usepackage[pdftex]{graphicx}
  \graphicspath{{../pdf/}{../jpeg/}}
  \DeclareGraphicsExtensions{.pdf,.jpeg,.png}
\else
  \usepackage[dvips]{graphicx}
  \graphicspath{{../eps/}}
  \DeclareGraphicsExtensions{.eps}
\fi
\usepackage[cmex10]{amsmath}
\usepackage {amssymb}
\usepackage{algorithmic}
\usepackage{array}
\usepackage{mdwmath}
\usepackage{mdwtab}
\usepackage{eqparbox}
\usepackage{url}
\usepackage{hyperref}
\usepackage{algorithm}
\usepackage{algorithmic}
\usepackage{epstopdf,cite,color}
\usepackage{amsthm} 

\usepackage{amsmath, amssymb, graphicx, booktabs, hyperref}

\renewcommand{\baselinestretch}{1}

\newcommand{\beq}{\begin{equation}}
\newcommand{\eeq}{\end{equation}}
\newcommand{\beqn}{\begin{eqnarray}}
\newcommand{\eeqn}{\end{eqnarray}}
\newcommand{\beqno}{\begin{eqnarray*}}
\newcommand{\eeqno}{\end{eqnarray*}}
\newcommand{\bma}{\begin{displaymath}}
\newcommand{\ema}{\end{displaymath}}
\newcommand{\bnu}{\begin{enumerate}}
\newcommand{\enu}{\end{enumerate}}
\newcommand{\bce}{\begin{center}}
\newcommand{\ece}{\end{center}}
\newcommand{\btb}{\begin{tabular}}
\newcommand{\etb}{\end{tabular}}

\begin{document}
\title{Quantum-Augmented AI/ML for O-RAN: Hierarchical Threat Detection with Synergistic Intelligence and Interpretability}

\author{Tan~Le,~\IEEEmembership{Member,~IEEE,} Van~Le, and Sachin~Shetty,~\IEEEmembership{Senior Member,~IEEE}
\thanks{T.~Le is with the School of Engineering, Architecture and Aviation, Hampton University, Hampton, VA 23669, USA. Email: tan.le@hamptonu.edu.}
\thanks{V.~Le is with the Virginia Polytechnic Institute and State University, Blacksburg, VA 24061, USA. Email: vanl@vt.edu.}
\thanks{S.~Shetty is with the Virginia Modeling, Analysis and Simulation Center, Old Dominion University, Suffolk, VA 23435, USA. 
Emails: sshetty@odu.edu.}}
\maketitle
\begin{abstract}
Open Radio Access Networks (O-RAN) enhance modularity and telemetry granularity but also widen the cybersecurity attack surface across disaggregated control, user, and management planes. We propose a hierarchical defense framework with three coordinated layers—anomaly detection, intrusion confirmation, and multiattack classification—each aligned with O-RAN’s telemetry stack. Our approach integrates hybrid quantum computing and machine learning, leveraging amplitude- and entanglement-based feature encodings with deep and ensemble classifiers. We conduct extensive benchmarking across synthetic and real-world telemetry, evaluating encoding depth, architectural variants, and diagnostic fidelity. The framework consistently achieves near-perfect accuracy, high recall, and strong class separability. Multi-faceted evaluation across decision boundaries, probabilistic margins, and latent space geometry confirms its interpretability, robustness, and readiness for slice-aware diagnostics and scalable deployment in near-RT and non-RT RIC domains.
\end{abstract}

\section{Introduction}

Open Radio Access Networks (O-RAN) offer modularity, interoperability, and vendor diversity for 5G deployments. However, this disaggregation introduces new vulnerabilities across control, user, and management planes \cite{11204489, 11045988, polese2023understanding, cisa2023, ntia2023}. Traditional flat security models—designed for monolithic architectures—fail to capture the layered escalation of threats in O-RAN environments, where subtle anomalies may propagate across slices before manifesting as confirmed intrusions or coordinated attacks \cite{de2023survey, analysysmason2025, 11204489, le2022artificial}.

Key challenges in O-RAN cybersecurity stem from architectural fragmentation, telemetry complexity, and model interpretability. First, the disaggregated nature of O-RAN interfaces—such as E2, A1, and O1—creates inconsistent security postures across vendors and domains, exposing control loops to cipher bidding-down attacks and partial encryption vulnerabilities \cite{10347507, groen2024securing, del2024cybersecurity, oranalliance2023, barker2025}. Second, the volume and heterogeneity of telemetry generated by SMO, near-RT RIC, and CU/DU layers overwhelm traditional detection pipelines, making it difficult to correlate anomalies in real time \cite{babar2025autonomous, ntia2023}. Third, classical machine learning models often lack interpretability and slice-awareness, limiting their utility in zero-trust architectures and policy-driven enforcement scenarios \cite{10812212, de2025quantum}.
Despite recent advances, several unsolved problems remain. Encoding high-dimensional telemetry for robust anomaly detection without sacrificing interpretability is still an open challenge. Confirming intrusions across fused telemetry streams while maintaining low false positive rates requires models that balance sensitivity and generalization. Finally, classifying multiattack scenarios with forensic traceability and cross-slice generalization remains difficult, especially under dynamic threat conditions and limited labeled data \cite{yang2024mismatched, singh2024, alqithami2025}.

To address these gaps, we introduce a hierarchical cybersecurity framework that mirrors the escalation of cyber threats. Our system integrates quantum-inspired feature encoding with hybrid model architectures to support slice-aware diagnostics, telemetry fusion, and forensic classification across three operational layers. This layered approach aligns with O-RAN’s telemetry stack and supports real-time, policy-driven security enforcement.

\textbf{Our contributions are:}
\begin{itemize}
    \item A three-layer defense stack aligned with O-RAN telemetry granularity—Layer 1 for anomaly detection, Layer 2 for intrusion confirmation, and Layer 3 for multiattack classification.
    \item Quantum-inspired encoding schemes—including amplitude-based and entanglement-based methods—that enhance feature separability and interpretability across hybrid quantum-classical pipelines.
    \item Benchmarking of Serial and Parallel Platforms for the Hybrid Quantum Computing and Deep Neural Networks (DQNN+DNN) as well as the Hybrid Quantum Machine Learning (QML) and Random Forest (RF) architectures across synthetic and real-world telemetry, demonstrating superior performance over classical baselines.
    \item Deployment validation through confusion matrix and t-SNE analyses, highlighting the framework’s suitability for toggled diagnostics and slice-specific traceability in near-RT and non-RT RIC domains.
\end{itemize}

\section{Related Work}

\textbf{Defense-in-depth for O-RAN:} The CISA and NTIA reports emphasize layered security integration across O-RAN interfaces, highlighting the need for telemetry-driven threat response and zero-trust enforcement \cite{cisa2023,ntia2023}. Analysys Mason’s 2025 study notes that operators still face uncertainty in de-risking O-RAN deployments due to fragmented security controls and lack of slice-specific diagnostics \cite{analysysmason2025}.

\textbf{Quantum-inspired encoding in cybersecurity:} NIST’s finalized post-quantum encryption standards \cite{nist2024} underscores the urgency of transitioning to quantum-safe systems. Quantum encoding techniques—such as amplitude encoding and entanglement-based transformations—enable richer latent representations and improved classification fidelity in hybrid models \cite{quantum_ai_dive, cao2019quantum, cao2017quantum}.

\textbf{Hierarchical reinforcement learning for cyber defense:} Recent work on hierarchical multi-agent reinforcement learning (MARL) demonstrates its potential for adaptive cyber response across threat escalation stages \cite{singh2024,alqithami2025,alshamrani2025, Tan18d, Tan2024, Tan18b}. However, these approaches often suffer from policy instability and require extensive tuning, limiting their deployment viability in real-time O-RAN environments.

Our framework builds on these foundations by integrating quantum-enhanced encoding with hierarchical model architectures, offering a scalable and interpretable solution for slice-aware cyber defense in O-RAN.

\section{System Model}

\subsection{O-RAN Architecture Integration}
\begin{figure*}[ht]
\centering
\includegraphics[width=.8\linewidth]{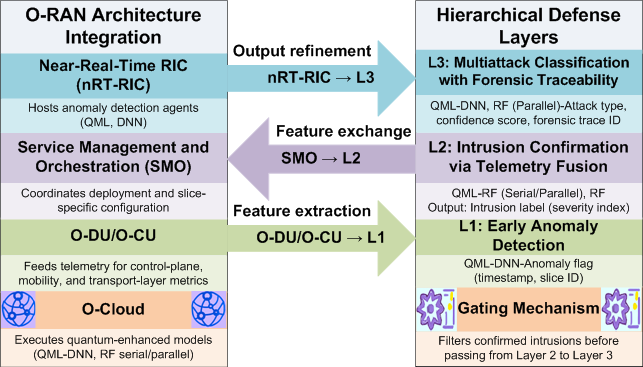}
\caption{System Model: Hierarchical Defense Framework Integration into O-RAN Architecture. The left section shows O-RAN components including nRT-RIC, SMO, O-Cloud, and O-DU/O-CU. The right section presents the defense layers: L1 (Anomaly Detection), L2 (Intrusion Confirmation), and L3 (Multiattack Classification), each with model type and output format. Arrows denote feature extraction, exchange, and output refinement.}
\label{fig:landscapecyber}
\end{figure*}
Figure~\ref{fig:landscapecyber} illustrates the integration of our hierarchical defense framework into the O-RAN architecture. The left half of the diagram shows the modular components of O-RAN—including the Near-Real-Time RIC (nRT-RIC), Service Management and Orchestration (SMO), O-DU/O-CU, and O-Cloud—while the right half presents the three-layer defense stack. Each layer is mapped to specific telemetry sources and model types, with arrows denoting feature extraction, exchange, and output refinement.

The framework is designed for modular deployment across virtualized O-RAN components. Layer 1 anomaly detection agents are hosted within the nRT-RIC, implemented as unsupervised modules with toggled quantum-inspired encoding and diagnostic logging. The SMO coordinates model deployment, version control, and slice-specific configuration across all layers. Quantum-enhanced models are executed in the O-Cloud, where toggles support encoding selection, ablation studies, and interpretability mapping. Telemetry streams from O-DU and O-CU provide control-plane signaling, mobility traces, and transport-layer metrics for feature extraction and cross-layer fusion.

\subsection{Defense Layers and Model Pipeline}

Table~\ref{tab:defense_layers} summarizes the functionality, model implementation, and output format for each layer in the hierarchical pipeline. Each layer is implemented as a modular code block with reproducible toggles, ablation-ready logic, and interpretable outputs.

Layer 1 performs early anomaly detection using slice-aware thresholds and toggled feature diagnostics, powered by quantum-inspired QML-DNN models. Layer 2 confirms intrusions through telemetry fusion and gating logic, using QML-RF and classical RF models to process control-plane and transport-layer features. Layer 3 performs multiattack classification with forensic traceability, leveraging engineered threat indicators and parallel QML-DNN and RF models for interpretable decision mapping.

\begin{table*}[h!]
\centering
\caption{Hierarchical Defense Layers and Model Implementation}
\begin{tabular}{|c|p{4.5cm}|p{4.5cm}|p{3.5cm}|}
\hline
\textbf{Layer} & \textbf{Function} & \textbf{Model Implementation} & \textbf{Output Format} \\
\hline
L1 & Early anomaly detection using slice-aware thresholds and toggled feature diagnostics & Quantum-inspired QML-DNN models with telemetry-driven encoding & Binary anomaly flag with timestamp and slice ID \\
\hline
L2 & Intrusion confirmation via telemetry fusion and gating logic & QML-RF (serial/parallel) and RF models leveraging control-plane and transport metrics & Intrusion label with severity index \\
\hline
L3 & Multiattack classification with forensic traceability and interpretability mapping & QML-DNN and RF (parallel) models using engineered threat indicators & Attack type label (0–5), confidence score, and forensic trace ID \\
\hline
\end{tabular}
\label{tab:defense_layers}
\end{table*}

\section{Threat Model}
\begin{figure*}[ht]
\centering
\vspace{-0.1in}
\mbox{\subfigure[]{\includegraphics[width=2.4in]{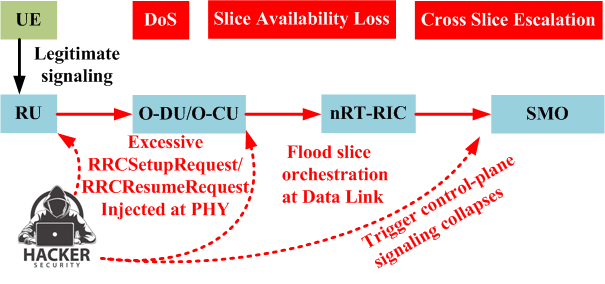}\label{Type1} }
\subfigure[]{\includegraphics[width=1.8in]{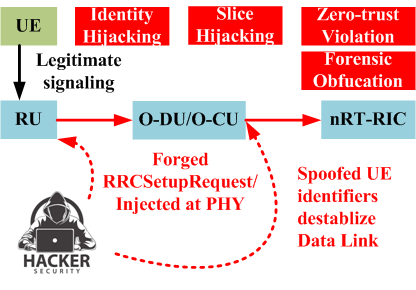}
\label{Type2}}
\subfigure[]{\includegraphics[width=2.4in]{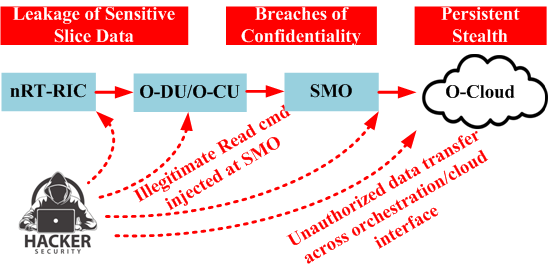} 
\label{Type3}}
}
\vspace{-0.1in}
\caption{(a) Type 1-Signaling Floods originate at the UE–PHY interface, where the adversary injects excessive RRCSetup/Resume requests into O‑RU/O‑DU. The attack then propagates upward through the Data Link Layer to the Near‑RT RIC and ultimately destabilizes SMO control at the Network Layer.
(b) Type 2-Spoofing begins at the UE–PHY interface, with adversaries presenting fake UE identities and falsified signaling to O‑RU/O‑DU. The attack flows upward into the Near‑RT RIC, where spoofing deceives admission control with falsified signaling, misallocating slice resources and destabilizing orchestration.
(c) Type 3-Exfiltration initiates at the SMO–O‑Cloud interface, where adversaries exploit orchestration–cloud communication. The attack direction is lateral across orchestration channels, enabling unauthorized data transfer and confidentiality compromise.}
\label{Threattype1}
\vspace{-.1in}
\end{figure*}

\begin{figure*}[ht]
\centering
\vspace{-0.1in}
\mbox{\subfigure[]{\includegraphics[width=2.4in]{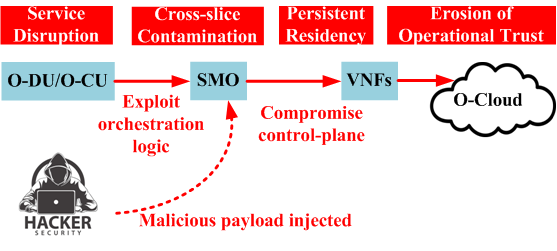} 
\label{Type4}}
\subfigure[]{\includegraphics[width=2.4in]{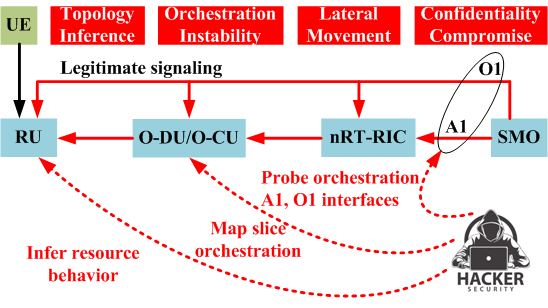} 
\label{Type5}}
}
\vspace{-0.1in}
\caption{(a) Type 4-Malware Injection starts within O‑Cloud workloads, where adversaries introduce malicious code. The attack direction is inward, compromising virtualized RAN functions and cascading upward to affect orchestration integrity and availability. (b) Type 5-Reconnaissance originates at A1/O1 interfaces at the Network Layer. The attack direction cascades downward into the Data Link and PHY layers, mapping slice orchestration and inferring resource behavior, which enables topology inference, lateral movement and confidentiality compromise.}
\label{Threattype2}
\vspace{-.2in}
\end{figure*}

Open RAN (O-RAN) introduces vulnerabilities that adversaries can exploit because of its disaggregated architecture, multi-vendor interfaces, and slice-aware orchestration. Unlike monolithic RANs, O-RAN exposes control, user, and management planes through standardized interfaces such as E2, A1, O1, and OFH. This openness creates opportunities for attackers to manipulate signaling, telemetry, and orchestration logic. We identify five distinct types of attacks—signaling floods, spoofing, exfiltration, malware and reconnaissance—and describe each in terms of adversary assumptions, attack vectors, and impact in the following.
Figs. \ref{Threattype1} and \ref{Threattype2} demonstrate these attack types.


\noindent\textbf{Type 1-Signaling Floods:}
In the case of signaling floods, the adversary assumptions are that attackers possess SDRs and open-source cellular stacks, giving them the capability to generate excessive signaling traffic before the security context is established. The attack vector involves injecting large volumes of RRCSetupRequest and RRCResumeRequest messages into the E2 interface, exploiting weak integrity protection in the RRC Inactive state. By overwhelming anomaly detectors and destabilizing slice orchestration, the attacker forces the system into overload. The impact is severe: legitimate UEs are denied service, slice availability is degraded, and resource contention spreads across slices, amplifying disruption throughout the network.

\noindent\textbf{Type 2-Spoofing:}
Spoofing attacks assume that adversaries can monitor and inject forged RRC/NAS messages with spoofed UE identifiers such as TMSI or SUCI. The attack vector here is the exploitation of weak integrity protection in early RRC signaling, allowing the attacker to impersonate legitimate UEs and bypass authentication. This grants unauthorized access to slice resources and undermines trust boundaries. The impact includes identity hijacking, unauthorized slice usage, and violations of zero-trust enforcement. Moreover, forensic analysis becomes complicated, as spoofed sessions obscure attribution and hinder traceability.

\noindent\textbf{Type 3-Exfiltration:}
For exfiltration, the adversary assumptions are that attackers can establish long-lived asymmetric flows that evade correlation across SMO, RIC, and O-Cloud telemetry. The attack vector involves manipulating control-plane signaling to mask data leakage, exploiting gaps in telemetry fusion and inconsistent monitoring across disaggregated components. These flows mimic legitimate slice behavior, making them difficult to detect. The impact is the leakage of sensitive slice data to external endpoints, breaches of confidentiality, and persistent stealth. Forensic traceability is lost, as the attacker maintains exfiltration channels across slice boundaries without triggering alarms.

\noindent\textbf{Type 4-Malware:}
Malware-based threats in O-RAN exploit vulnerabilities in virtualized network functions (VNFs) and orchestration pipelines to embed malicious payloads that persist across slices. Adversaries assume access to misconfigured deployment environments or compromised supply chain artifacts, enabling them to inject code into VNFs hosted in the O-Cloud. Attack vectors include container escape, hypervisor exploitation, and manipulation of orchestration APIs (e.g., O1, A1) to escalate privileges and propagate malware across slices. Once embedded, the payload can hijack VNFs, corrupt slice orchestration, and manipulate control-plane logic. The resulting impact includes service disruption, cross-slice contamination, persistent malware residency through virtualization, and erosion of operational trust across the O-RAN architecture


\noindent\textbf{Type 5-Reconnaissance:}
Reconnaissance attacks assume that adversaries can probe orchestration interfaces such as A1 and O1 without raising suspicion. The attack vector consists of systematically querying slice orchestration metrics to infer resource allocation patterns and slice priorities. By mapping slice topology and identifying high-value targets, the attacker prepares for more disruptive operations. The impact is strategic: resource mapping enables targeted disruption, undermines orchestration stability, and facilitates lateral movement across slices. Confidentiality of slice scheduling behavior is compromised, leaving the system vulnerable to subsequent attacks.

\noindent\textbf{Defense Framework:}
To mitigate these threats, we propose a hierarchical three-layer defense framework that aligns detection granularity with O-RAN’s telemetry stack. The first layer focuses on anomaly detection, identifying deviations in signaling rates, flow symmetry, and telemetry access. The second layer escalates suspicious events to intrusion confirmation using quantum-augmented deep quantum neural networks, which remain resilient under adversarial noise and telemetry poisoning. The third layer provides multiattack classification with slice-aware interpretability, distinguishing signaling floods, spoofing, exfiltration, and reconnaissance with high fidelity. Together, these layers restore zero-trust enforcement, enable forensic traceability, and adapt to the operational realities of O-RAN deployments.

We summarize the threat model and our defense strategy as follows.

\subsection{Adversary Assumptions}
We assume a capable adversary equipped with commercial SDRs and open-source cellular stacks, consistent with Det-RAN~\cite{scalingi2024detran} and 5G-SPECTOR~\cite{wen2024specter}. Such adversaries can:
\begin{itemize}
    \item Inject forged RRC/NAS messages prior to security context establishment, exploiting unprotected L3 signaling.
    \item Replay or spoof UE identities (e.g., TMSI, SUCI) to trigger denial-of-service or unauthorized slice access.
    \item Poison telemetry streams across SMO, nRT-RIC, and O-Cloud to mask malicious activity.
    \item Exploit weak encryption or inconsistent enforcement across E2, A1, and O1 interfaces to bypass zero-trust enforcement.
\end{itemize}

\subsection{O-RAN-Specific Threats}
Building on recent threat modeling frameworks~\cite{dessources2025oranrisk, ieee2024formal}, we identify five primary O-RAN-specific exploits:
\begin{itemize}
    \item \textbf{Signaling floods:} Excessive \texttt{RRCSetupRequest} or \texttt{RRCResumeRequest} messages overwhelm nRT-RIC anomaly detectors, degrading slice availability and destabilizing admission control.
    \item \textbf{Spoofing:} Weak integrity protection in the RRC Inactive state allows adversaries to impersonate legitimate UEs and inject forged identifiers, enabling unauthorized slice access and mobility disruption.
    \item \textbf{Exfiltration:} Long-lived asymmetric flows bypass telemetry fusion across SMO, RIC, and O-Cloud, leaking sensitive slice data and violating confidentiality and forensic traceability.
    \item \textbf{Malware:} Malicious payloads exploit vulnerabilities in virtualized network functions (VNFs). Malware is injected via compromised orchestration pipelines, vulnerable APIs, or lateral movement across slices. Once embedded in VNFs hosted in the O-Cloud, the payload hijacks orchestration logic and control-plane behavior, enabling persistent residency, cross-slice infection, and long-term service disruption.
    \item \textbf{Reconnaissance:} Probing slice orchestration metrics via A1 and O1 interfaces reveals resource allocation patterns, enabling topology inference, targeted disruption, and lateral movement across slices.
\end{itemize}

\subsection{Impact}
These threats degrade slice availability, compromise forensic traceability, and undermine zero-trust enforcement. By exploiting O-RAN’s disaggregation, adversaries achieve stealth and persistence not possible in monolithic RANs. Our hierarchical defense framework directly addresses these scenarios by aligning anomaly detection, intrusion confirmation, and multiattack classification with O-RAN's telemetry stack.

So now, we present how adversaries compromise the O-RAN system by exploiting its disaggregated architecture and standardized interfaces. 
Attackers initiate \textbf{signaling floods} by injecting excessive \texttt{RRCSetupRequest} or \texttt{RRCResumeRequest} messages before security context establishment, overwhelming nRT-RIC anomaly detectors and degrading slice availability. 
\textbf{Spoofing attacks} exploit weak integrity protection in the RRC Inactive state, allowing adversaries to impersonate legitimate UEs and gain unauthorized slice access. 
\textbf{Exfiltration} occurs when long-lived asymmetric flows bypass telemetry fusion across SMO, RIC, and O-Cloud, leaking sensitive slice data and violating confidentiality. 
\textbf{Malware attacks} exploit vulnerabilities in virtualized network functions (VNFs) hosted in the O-Cloud. Adversaries inject malicious payloads via compromised orchestration pipelines, vulnerable APIs, or lateral movement across slices. Once embedded, the malware hijacks orchestration logic and control-plane behavior, enabling persistent residency, cross-slice infection, and long-term service disruption. 
\textbf{Reconnaissance attacks} probe orchestration metrics via A1 and O1 interfaces to infer resource allocation patterns, enabling topology inference, targeted disruption, and lateral movement.

These threats are amplified by O-RAN's modular design, where inconsistencies across E2, A1, and O1 interfaces allow adversaries to persist across slices and evade detection. 
Our proposed \textbf{three-layer defense framework} counters these threats by aligning detection logic with O-RAN’s telemetry stack. 
\textbf{Layer 1} performs lightweight anomaly detection using fused control-plane and user-plane telemetry, flagging deviations in signaling rates, flow symmetry, and interface access. 
\textbf{Layer 2} escalates flagged events to binary intrusion confirmation using quantum-augmented deep quantum neural network (DQNN) classifiers, which remain resilient under adversarial noise and telemetry poisoning. 
\textbf{Layer 3} performs multiattack classification with slice-aware interpretability, leveraging entanglement-enhanced latent embeddings to distinguish signaling floods, spoofing, exfiltration, malware, and reconnaissance with high fidelity. 
This hierarchical approach restores zero-trust enforcement, enables forensic traceability, and adapts to the operational realities of O-RAN deployments.

\section{Semantic Flow and Feature Progression Across Detection Layers}

\subsection{Hierarchical Cybersecurity Framework: Layer Relationships}

\begin{figure}[h]
\centering
\includegraphics[width=0.5\linewidth]{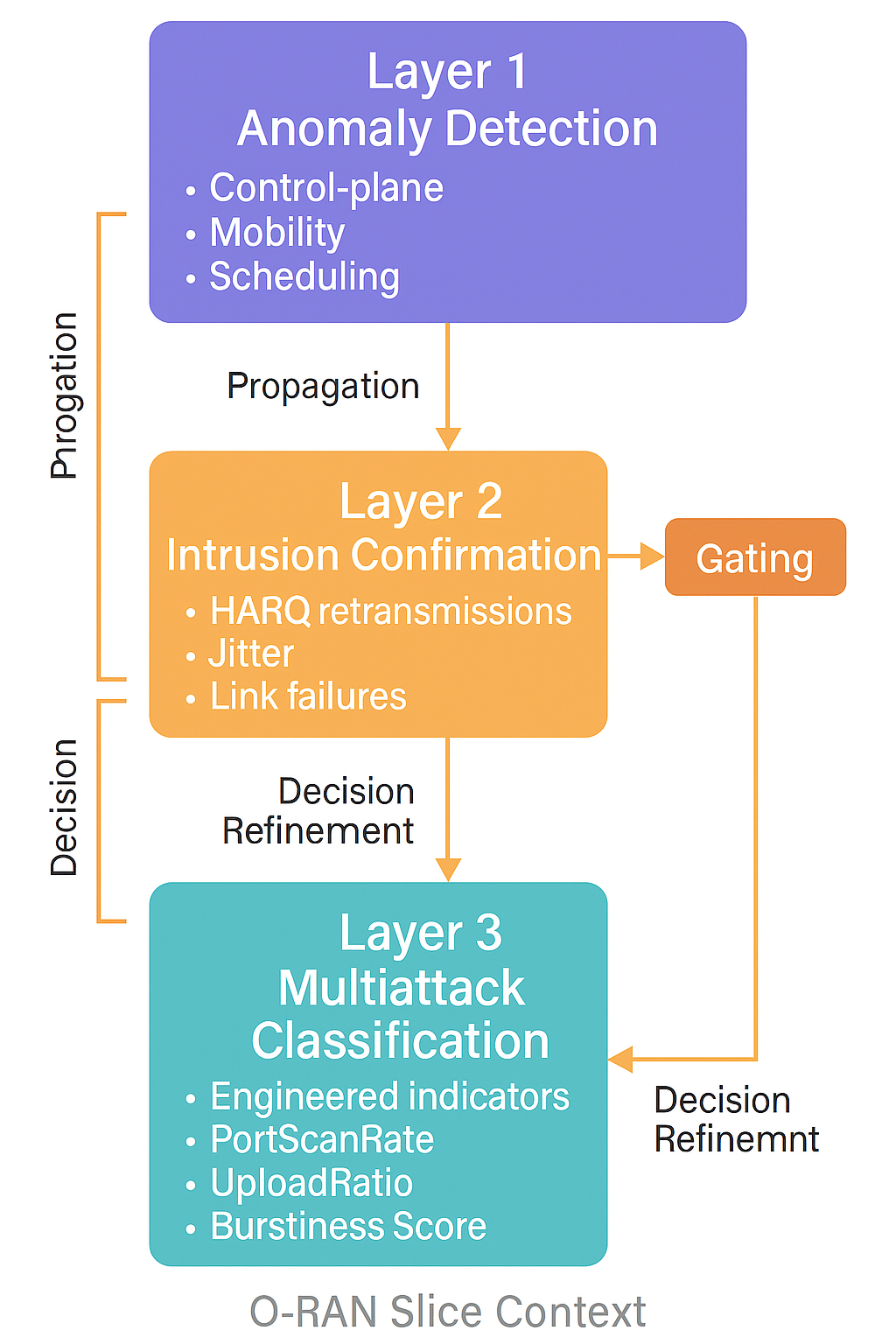}
\caption{Hierarchical cybersecurity framework for O-RAN slice-aware threat detection. \textbf{Layer 1} detects early anomalies using control-plane and mobility metrics. \textbf{Layer 2} confirms intrusions via low-level telemetry such as HARQ retransmissions and jitter. A dedicated \textbf{gating mechanism} filters confirmed threats before activating \textbf{Layer 3}, which classifies attack types using engineered indicators like PortScanRate and UploadRatio. Arrows denote decision refinement and feature propagation across layers.}
\label{fig:hierarchical_framework_transition}
\end{figure}

To illustrate the overall structure and flow of the proposed framework, Figure~\ref{fig:hierarchical_framework_transition} presents a high-level view of the three-layered architecture, including the gating mechanism and feature propagation logic.

The proposed framework adopts a three-layered architecture for interpretable, slice-aware threat detection in O-RAN environments. Each layer contributes a distinct function within a defense-in-depth strategy:

\begin{itemize}
    \item \textbf{Layer 1: Anomaly Detection} — Detects early signs of abnormal behavior using control-plane and mobility metrics.
    \item \textbf{Layer 2: Intrusion Confirmation} — Validates whether anomalies reflect true intrusions using low-level telemetry.
    \item \textbf{Layer 3: Multiattack Classification} — Diagnoses and labels specific cyberattack types using engineered threat indicators.
\end{itemize}

\subsubsection*{Layered Flow Summary}

\begin{center}
\texttt{Layer 1} $\rightarrow$ \texttt{Layer 2} $\rightarrow$ \texttt{Layer 3}
\end{center}

This layered structure supports scalable, interpretable, and modular threat detection across virtualized O-RAN slices \cite{basaran2025xainomaly, unicorn2025, oranThreatModel2025}.

\subsection{Cross-Layer Feature Logic and Progression}

The hierarchical cybersecurity framework integrates three layers of detection, each operating on distinct feature sets. These features are designed to be complementary and logically connected, enabling traceable decision refinement across the O-RAN threat detection pipeline.

\subsubsection*{Layer Roles and Feature Scope}

Each layer in the framework operates on distinct feature sets and contributes to progressively refined threat detection:

\begin{itemize}
    \item \textbf{Layer 1: Anomaly Detection} — Monitors control-plane signaling, mobility patterns, and scheduling anomalies to detect early deviations from expected behavior.
    \item \textbf{Layer 2: Intrusion Confirmation} — Validates anomalies using physical-layer and transport-level telemetry such as HARQ retransmissions, jitter, and registration failures.
    \item \textbf{Layer 3: Multiattack Classification} — Diagnoses and labels confirmed intrusions using engineered threat indicators including PortScanRate, UploadRatio, and Burstiness Score.
\end{itemize}

\subsubsection*{Inter-Layer Feature Relationships (Table~\ref{tab:cross_layer_feature_mapping})}

Table~\ref{tab:cross_layer_feature_mapping} summarizes how features propagate across layers. Each Layer 1 anomaly is validated by a corresponding Layer 2 metric and mapped to one or more Layer 3 classification features.

\begin{table*}[h]
\centering
\caption{Logical Relationships Between Input Features Across Layers}
\begin{tabular}{|p{4cm}|p{4.5cm}|p{4.5cm}|}
\hline
\textbf{Layer 1 Feature} & \textbf{Validated by Layer 2 Feature} & \textbf{Mapped to Layer 3 Feature} \\
\hline
Connection Request Rate & Registration Failure Rate & UnauthorizedAccess, PortScanRate \\
\hline
Connection Setup Time & HARQ Retransmission Count & DoS Burstiness Score, QoS Violation Frequency \\
\hline
Mobility Index & Radio Link Failure Rate & Spoofing Signal Deviation \\
\hline
Paging Response Rate & Registration Failure Rate & Replay Timing Offset \\
\hline
Control Plane Entropy & Packet Size Variance & MaliciousPayloadSize, Slice Resource Deviation \\
\hline
Scheduling Anomaly Score & Jitter Index & UploadRatio, Slice Resource Deviation \\
\hline
\end{tabular}
\label{tab:cross_layer_feature_mapping}
\end{table*}

\subsubsection*{Interpretability Summary}

This feature-level mapping enables:

\begin{itemize}
    \item \textbf{Traceable decision refinement} — Each classification is grounded in validated anomalies.
    \item \textbf{Modular feature design} — Features are aligned with O-RAN control, user, and management planes.
    \item \textbf{Explainable threat detection} — Supports forensic analysis and targeted response.
\end{itemize}

\subsubsection*{Layer Transition Integrity}

The transitions between layers in the proposed framework are designed to be logically coherent, operationally efficient, and semantically aligned. Moving from \textbf{Layer 1: Anomaly Detection} to \textbf{Layer 2: Intrusion Confirmation} presents no inconvenience. The output of Layer 1—such as anomaly flags based on signaling, mobility, or scheduling irregularities—is directly traceable to Layer 2 inputs like HARQ retransmission count, registration failure rate, and jitter index. These Layer 2 features validate whether the detected anomalies reflect true intrusions, using deeper physical-layer and transport-level evidence. The semantic alignment between these layers is strong, and the telemetry required for Layer 2 is typically available within the same monitoring infrastructure, ensuring computational feasibility.

Transitioning from \textbf{Layer 2} to \textbf{Layer 3: Multiattack Classification} also follows a clear logic, but warrants a minor operational caution. While Layer 2 confirms the presence of an intrusion, the activation of Layer 3 must be gated to ensure that classification is performed only on validated threats. This prevents premature or erroneous labeling of benign anomalies. Once Layer 2 confirms an intrusion, Layer 3 can be triggered to specify the exact attack type—such as DoS, Spoofing, or Exfiltration—using engineered threat indicators. This classification not only supports real-time mitigation but also enables forensic applications such as incident reconstruction, threat attribution, and slice-level impact analysis. By ensuring that Layer 3 operates only on confirmed intrusions, the framework preserves semantic clarity, reduces false positives, and enhances the interpretability of downstream decisions.

To further reinforce this transition, a \textbf{gating mechanism} should be implemented, whereby Layer 3 classifiers are activated only when Layer 2 outputs exceed a predefined confidence threshold or anomaly severity index. Additionally, \textbf{temporal linking} between layers—such as enforcing a maximum delay between Layer 1 anomaly detection and Layer 2 confirmation—can improve responsiveness and reduce ambiguity. These enhancements ensure that the hierarchical pipeline remains robust, explainable, and suitable for both operational defense and post-incident forensic analysis.

Overall, the layered transitions support \textbf{progressive abstraction}, \textbf{cross-layer traceability}, and \textbf{modular interpretability}. Each decision in Layer 3 can be traced back to validated evidence in Layer 2 and anomaly triggers in Layer 1. This structure reinforces defense-in-depth and ensures that threat detection across O-RAN slices remains scalable, explainable, and operationally sound.

\subsection{Layer 1: Anomaly Detection in O-RAN}
\label{AnomalyORAN}
\subsubsection*{Input Features}
Layer 1 monitors control-plane and mobility-related metrics to detect early signs of anomalous behavior across O-RAN slices. These features are extracted from UE registration, signaling, and scheduling telemetry, enabling proactive detection of service-level disruptions \cite{oranThreatModel2025, unicorn2025}:

\begin{itemize}
    \item \textbf{Connection Request Rate} — Frequency of RRC connection requests per UE. Sudden spikes may indicate signaling floods or DoS attempts.

    \item \textbf{Connection Setup Time} — Average time to complete RRC setup. Prolonged setup durations may reflect congestion, spoofing, or control-plane interference.

    \item \textbf{Mobility Index} — UE handover frequency and distance. Abnormal mobility patterns may signal spoofing, rogue cell behavior, or reconnaissance.

    \item \textbf{Paging Response Rate} — Ratio of successful paging responses. Low response rates may indicate UE unavailability due to jamming, spoofing, or registration failure.

    \item \textbf{Control Plane Entropy} — Entropy of signaling message types. Elevated entropy may reflect protocol misuse, scanning, or malware activity.

    \item \textbf{Scheduling Anomaly Score} — Deviation from expected scheduling patterns across slices. Irregular scheduling may result from resource contention or adversarial manipulation.
\end{itemize}

\subsubsection*{Attack Type Relevance}
These features are designed to detect early-stage anomalies that may precede confirmed intrusions. Layer 1 acts as a proactive filter, flagging suspicious behaviors for deeper inspection by downstream layers.

\begin{table*}[h]
\centering
\caption{Mapping of Layer 1 Features to Behavioral Signatures and Associated Attack Types}
\begin{tabular}{|p{3.5cm}|p{5cm}|p{5.5cm}|}
\hline
\textbf{Layer 1 Feature} & \textbf{Behavioral Signature} & \textbf{Associated Attack Types} \\
\hline
\textbf{Connection Request Rate} & Signaling floods or excessive UE registration attempts & DoS, Spoofing \\
\hline
\textbf{Connection Setup Time} & Prolonged setup due to congestion or signaling interference & DoS, Malware, Spoofing \\
\hline
\textbf{Mobility Index} & Abnormal handover frequency or distance & Spoofing, Reconnaissance \\
\hline
\textbf{Paging Response Rate} & Low response rates due to jamming or spoofed UE states & Spoofing, DoS \\
\hline
\textbf{Control Plane Entropy} & High entropy from protocol misuse or scanning activity & Malware, Reconnaissance \\
\hline
\textbf{Scheduling Anomaly Score} & Irregular scheduling patterns across slices & DoS, Exfiltration \\
\hline
\end{tabular}
\label{tab:layer1_feature_mapping}
\end{table*}

\subsubsection*{Feature Detection Summary}
Layer 1 features serve as early anomaly indicators, enabling proactive detection of slice-aware disruptions before full intrusion confirmation. This supports defense-in-depth by initiating multi-layered threat detection across the O-RAN stack \cite{basaran2025xainomaly}.

\subsection{Layer 2: Intrusion Confirmation in O-RAN}

\subsubsection*{Input Features}
Layer 2 uses six low-level metrics derived from UE telemetry and slice-level fault indicators to validate potential intrusions flagged by upstream classifiers. These features reflect physical-layer reliability, transport-layer timing, and control-plane integrity across the O-RAN stack \cite{oranThreatModel2025, unicorn2025}:

\begin{itemize}
    \item \textbf{HARQ Retransmission Count} — Number of Hybrid Automatic Repeat Request retransmissions per UE. Elevated counts may indicate link-layer congestion, interference, or jamming.
    
    \item \textbf{Flow Interarrival Time} — Time between consecutive packet flows. Irregular timing patterns may signal replay attacks, DoS bursts, or stealthy exfiltration.
    
    \item \textbf{Packet Size Variance} — Variability in packet sizes across flows. High variance may reflect mixed payloads typical of malware or reconnaissance probes.
    
    \item \textbf{Jitter Index} — Normalized jitter across flows, computed from inter-packet arrival deviations. Elevated jitter may indicate QoS degradation or slice-level contention.
    
    \item \textbf{Radio Link Failure Rate} — Frequency of radio link failures per UE. Persistent failures may result from spoofing, rogue cell interference, or targeted disruption.
    
    \item \textbf{Registration Failure Rate} — Ratio of failed UE registration attempts. Anomalous spikes may reflect identity spoofing or control-plane denial-of-service.
\end{itemize}

\subsubsection*{Attack Type Relevance}
These features are designed to confirm slice-aware anomalies associated with specific attack types. While Layer 3 performs multiattack classification, Layer 2 validates whether observed behaviors are consistent with physical or control-plane disruptions.

\begin{table*}[h]
\centering
\caption{Mapping of Layer 2 Features to Behavioral Signatures and Associated Attack Types}
\begin{tabular}{|p{3.5cm}|p{5cm}|p{5.5cm}|}
\hline
\textbf{Layer 2 Feature} & \textbf{Behavioral Signature} & \textbf{Associated Attack Types} \\
\hline
\textbf{HARQ Retransmission Count} & Link-layer congestion or interference due to bursty or spoofed traffic & DoS, Spoofing, Malware \\
\hline
\textbf{Flow Interarrival Time} & Irregular timing patterns or replayed flows & Replay, Exfiltration, DoS \\
\hline
\textbf{Packet Size Variance} & Mixed payloads or scanning probes across flows & Malware, Reconnaissance \\
\hline
\textbf{Jitter Index} & QoS degradation from slice contention or malicious interference & DoS, Exfiltration \\
\hline
\textbf{Radio Link Failure Rate} & Persistent link instability or rogue signal injection & Spoofing, DoS \\
\hline
\textbf{Registration Failure Rate} & Failed UE registrations due to identity spoofing or signaling disruption & Spoofing, Malware \\
\hline
\end{tabular}
\label{tab:layer2_feature_mapping}
\end{table*}

\subsubsection*{Feature Validation Summary}
Layer 2 features serve as intrusion confirmation signals, validating whether slice-level anomalies are consistent with physical-layer disruptions or control-plane compromise. This supports defense-in-depth by reinforcing Layer 3 decisions with low-level evidence \cite{basaran2025xainomaly}.

\subsection{Layer 3: Multi-attack Classification in O-RAN}

\subsubsection*{Input Features}
Each sample is represented by an 11-dimensional feature vector derived from O-RAN performance metrics and slice-level security indicators. These features are engineered to reflect realistic cyber-physical behaviors across disaggregated network functions and service slices \cite{oranThreatModel2025, unicorn2025}:

\begin{itemize}
    \item \textbf{Network KPIs (5 features):}
    \begin{itemize}
        \item \textbf{RSRP} — Reference Signal Received Power from distributed units (DUs)
        \item \textbf{RSRQ} — Reference Signal Received Quality across radio resource blocks
        \item \textbf{RSSI-NR} — Received Signal Strength Indicator for New Radio (NR) links
        \item \textbf{Throughput} — Aggregate data rate across service slices
        \item \textbf{PRB Usage} — Physical Resource Block utilization per slice
    \end{itemize}
    \item \textbf{Security Indicators (4 features):}
    \begin{itemize}
        \item \textbf{PortScanRate} — Frequency of port scanning activity targeting O-RAN interfaces
        \item \textbf{PacketDropRate} — Rate of dropped packets across midhaul and fronthaul segments
        \item \textbf{UnauthorizedAccess} — Binary flag for spoofing or malware attempts on control-plane functions
        \item \textbf{MaliciousPayloadSize} — Size of detected malicious content in user-plane traffic
    \end{itemize}
    \item \textbf{Exfiltration-Specific Features (2 features):}
    \begin{itemize}
        \item \textbf{ExfilFlowDuration} — Duration of long-lived flows indicative of stealthy data leakage
        \item \textbf{UploadRatio} — Ratio of upstream to downstream traffic across slices
    \end{itemize}
\end{itemize}

\subsubsection*{Attack Type Labels}
The dataset simulates six traffic categories, including one benign class and five adversarial types. Each attack type reflects a distinct threat scenario targeting O-RAN components and slice-level services \cite{zhang2024hierarchical}:

\begin{itemize}
    \item \textbf{Class 0: Normal} — Legitimate traffic with no malicious indicators.
    \item \textbf{Class 1: Denial-of-Service (DoS)} — Burst-based flooding of slice resources to degrade service availability.
    \item \textbf{Class 2: Spoofing} — Identity manipulation via forged signals or control-plane impersonation.
    \item \textbf{Class 3: Data Exfiltration} — Stealthy, asymmetric flows used to leak sensitive data from slices.
    \item \textbf{Class 4: Malware} — Malicious payloads exploiting vulnerabilities in virtualized network functions.
    \item \textbf{Class 5: Reconnaissance} — Scanning and probing behaviors targeting slice orchestration and topology.
\end{itemize}

\subsubsection*{Interpretable Classification Features}
To support explainable multiattack classification, Layer 3 integrates five interpretable indicators derived from slice-level metrics and temporal traffic patterns. These features are designed to align with the modular, service-oriented nature of O-RAN and support AI-driven threat detection with human-interpretable outputs \cite{basaran2025xainomaly, unicorn2025, oranThreatModel2025}:

\begin{itemize}
    \item \textbf{DoS Burstiness Score} — Quantifies the intensity and irregularity of traffic bursts targeting slice resources. DoS attacks often manifest as high-variance, short-duration spikes in throughput or PRB usage \cite{oranThreatModel2025}.
    
    \item \textbf{Spoofing Signal Deviation} — Measures deviations in signal characteristics such as RSRP and RSRQ that may indicate identity spoofing. In O-RAN, spoofing can compromise control-plane integrity by impersonating legitimate UEs or xApps \cite{rogueCell2025}.
    
    \item \textbf{Replay Timing Offset} — Captures temporal inconsistencies in packet sequences, such as unnatural interarrival times or duplicated flows. Replay attacks exploit timing gaps to inject stale or malicious traffic into slice-specific sessions \cite{basaran2025xainomaly}.
    
    \item \textbf{QoS Violation Frequency} — Tracks the rate of service-level agreement (SLA) breaches across slices. Attacks such as malware or exfiltration may degrade latency, jitter, or throughput, violating URLLC or eMBB slice constraints \cite{unicorn2025}.
    
    \item \textbf{Slice Resource Deviation} — Detects anomalies in resource allocation metrics (e.g., PRB usage, throughput) within individual slices. Reconnaissance or exfiltration may cause subtle shifts in resource distribution \cite{oranThreatModel2025}.
\end{itemize}

\begin{table*}[h]
\centering
\caption{Mapping of Attack Types to Behavioral Signatures and Interpretable Features}
\begin{tabular}{|p{3cm}|p{5.5cm}|p{5.5cm}|}
\hline
\textbf{Attack Type} & \textbf{Behavioral Signature} & \textbf{Relevant Interpretable Features} \\
\hline
\textbf{Normal} & Baseline traffic with no anomalies; conforms to expected slice-level KPIs and security indicators. & None (used for contrast and baseline calibration) \\
\hline
\textbf{Denial-of-Service (DoS)} & High-volume, bursty traffic patterns targeting slice throughput and PRB resources. & DoS Burstiness Score, QoS Violation Frequency, Slice Resource Deviation \\
\hline
\textbf{Spoofing} & Signal-level anomalies and identity inconsistencies; forged RSRP/RSRQ profiles. & Spoofing Signal Deviation, UnauthorizedAccess \\
\hline
\textbf{Data Exfiltration} & Long-lived, asymmetric flows with elevated upstream traffic; stealthy behavior. & ExfilFlowDuration, UploadRatio, QoS Violation Frequency \\
\hline
\textbf{Malware} & Malicious payloads and unauthorized access attempts; potential service degradation. & MaliciousPayloadSize, UnauthorizedAccess, QoS Violation Frequency \\
\hline
\textbf{Reconnaissance} & Scanning and probing of slice resources; subtle shifts in resource usage. & PortScanRate, Slice Resource Deviation \\
\hline
\end{tabular}
\label{tab:attack_feature_mapping}
\end{table*}

\subsubsection*{Feature Generation Summary}
The labeled dataset includes 10,000 samples across six classes (0–5), simulating both benign and adversarial traffic in a virtualized 5G O-RAN environment. Features are normalized and engineered to reflect slice-level behaviors and attack-specific perturbations, enabling interpretable and reproducible benchmarking for AI-driven threat detection.

\section{Hierarchical Multiattack Detection Formulation}
\begin{table*}[ht]
\centering
\caption{Summary of hierarchical multiattack detection layers, objectives, models, and simulators.}
\label{tab:hierarchy_summary}
\begin{tabular}{|c|p{4cm}|p{4cm}|p{3cm}|}
\hline
\textbf{Layer} & \textbf{Objective} & \textbf{Model(s)} & \textbf{Simulator} \\
\hline
1: Anomaly Detection &
Binary classification loss (cross-entropy) + latency penalty &
Quantum-inspired DQNN or classical DNN with toggled encoding depth and slice masking &
Qiskit Aer (local) \\
\hline
2: Intrusion Confirmation &
Binary classification loss (hinge or cross-entropy) + latency penalty under gating &
QML-RF (serial/parallel) or classical RF with telemetry fusion and gating logic &
IBM Q / NSF ACCESS \\
\hline
3: Multiattack Classification &
Multi-class loss (cross-entropy or macro-F1) + latency penalty + interpretability cost &
QML-DNN or parallel RF with engineered threat indicators and forensic trace ID logic &
IBM Q backend (noise mitigation) \\
\hline
\end{tabular}
\end{table*}
We formalize multiattack detection in 5G Open Radio Access Networks (O-RAN) as
a hierarchical classification pipeline $f = f_3 \circ f_2 \circ f_1$ operating
over cyber-physical telemetry. Each sub-function $f_\ell$ corresponds to a
defense layer $\ell \in \{1,2,3\}$:

\begin{itemize}
    \item $f_1$: anomaly detection in \textbf{nRT-RIC}
    \item $f_2$: intrusion confirmation in \textbf{O-Cloud} with SMO telemetry fusion
    \item $f_3$: multiattack classification in \textbf{O-Cloud} with forensic traceability
\end{itemize}

Let $\mathbf{x} \in \mathbb{R}^d$ denote a feature vector composed of
physical-layer KPIs (RSRP, RSRQ, RSSI-NR), network-layer metrics (PRB usage,
throughput), and cybersecurity indicators (port scan rate, packet drop rate,
unauthorized access flags, malicious payload size). The output space is
$\mathcal{Y} = \{0,1,2,3,4,5\}$, where 0 is \textbf{Normal}, 1 is
\textbf{DoS}, 2 is \textbf{Spoofing}, 3 is \textbf{Data Exfiltration}, 4 is
\textbf{Malware Injection}, and 5 is \textbf{Reconnaissance}.

\subsection{Unified Optimization Objective}
We define a general training objective across layers:
\beqn
\min_{f} \quad
\sum_{\ell=1}^3 \Big( \mathcal{L}_\ell(f_\ell(\mathbf{x}), y_\ell)
+ \lambda_\ell \cdot T_{\mathrm{inf}}(f_\ell) \Big)
+ \lambda_4 \cdot \mathcal{I}(f_3),
\eeqn
where:
\begin{itemize}
    \item $\mathcal{L}_\ell$: loss function for layer $\ell$
    \item $y_\ell$: ground-truth label (binary for $f_1,f_2$, multi-class for $f_3$)
    \item $T_{\mathrm{inf}}(f_\ell)$: inference latency under O-RAN constraints
    \item $\mathcal{I}(f_3)$: interpretability cost (e.g., SHAP entropy, model depth)
    \item $\lambda_\ell, \lambda_4$: trade-off hyperparameters
\end{itemize}

This unified form avoids repetition while capturing accuracy–latency–interpretability trade-offs.

\subsection{Layer-Specific Implementations}

\paragraph{Layer 1: Early Anomaly Detection}
\begin{itemize}
    \item \textbf{Objective:} Binary classification loss (cross-entropy) + latency penalty.
    \item \textbf{Model:} Quantum-inspired DQNN or classical DNN with toggled encoding depth and slice masking.
    \item \textbf{Simulator:} Local Qiskit Aer for fast toggling and ablation studies.
\end{itemize}

\paragraph{Layer 2: Intrusion Confirmation}
\begin{itemize}
    \item \textbf{Objective:} Binary classification loss (hinge or cross-entropy) + latency penalty under gating.
    \item \textbf{Model:} QML-RF (serial/parallel) or classical RF with telemetry fusion and gating logic.
    \item \textbf{Simulator:} IBM Q or NSF ACCESS for scalable inference.
\end{itemize}

\paragraph{Layer 3: Multiattack Classification}
\begin{itemize}
    \item \textbf{Objective:} Multi-class loss (cross-entropy or macro-F1) + latency penalty + interpretability cost.
    \item \textbf{Model:} QML-DNN or RF (parallel) with engineered threat indicators and forensic trace ID logic.
    \item \textbf{Simulator:} IBM Q backend with noise mitigation.
\end{itemize}

\subsection{Gated Prediction Pipeline}
The full prediction function $f(\mathbf{x})$ is gated across layers:
\[
f(\mathbf{x}) =
\begin{cases}
0 & \text{if } f_1(\mathbf{x}) = 0 \\
0 & \text{if } f_1(\mathbf{x}) = 1 \text{ and } f_2(\mathbf{x}) = 0 \\
f_3(\mathbf{x}) & \text{if } f_1(\mathbf{x}) = 1 \text{ and } f_2(\mathbf{x}) = 1
\end{cases}
\]

This hierarchical optimization framework enables reproducible benchmarking of
hybrid quantum-classical models against classical ML baselines, while
respecting O-RAN constraints on latency, explainability, and forensic
traceability.

\begin{table*}[ht]
\centering
\caption{Comparison of Encoding Strategies Across Layers}
\begin{tabular}{|p{3.5cm}|p{4cm}|p{4cm}|}
\hline
\textbf{Attribute} & \textbf{Amplitude Encoding} & \textbf{Fully Entangled Encoding} \\
\hline
Qubit Interaction & None (independent) & Global (entangled) \\
\hline
Circuit Depth & Shallow & Moderate to deep \\
\hline
Simulator & Qiskit Aer (local) & IBM Q / NSF ACCESS \\
\hline
Deployment Layer & L1 (Anomaly Detection) & L3 (Multiattack Classification) \\
\hline
Model Type & DQNN $\rightarrow$ DNN (serial) & QML-DNN or RF (parallel) \\
\hline
Use Case & Fast toggled diagnostics & Forensic traceability and multi-class separation \\
\hline
\end{tabular}
\label{tab:encoding_comparison}
\end{table*}

\begin{table*}[ht]
\centering
\caption{Benchmark Comparison of QML-RF and QML-DNN for O-RAN Layered Threat Detection}
\begin{tabular}{|p{3.5cm}|p{5.5cm}|p{5.5cm}|}
\hline
\textbf{Criterion} & \textbf{QML-RF (Layer 2)} & \textbf{QML-DNN (Layer 3)} \\
\hline
Gradient Sensitivity & Robust to gradient shifts due to tree-based logic; stable under noisy telemetry & Sensitive to gradient instability; performance degrades with non-stationary inputs \\
\hline
Interpretability & High interpretability via feature importance and gating traceability & Requires post-hoc explanation (e.g., SHAP, LRP) for forensic traceability \\
\hline
Simulator Compatibility & Efficient on IBM Q and NSF ACCESS; supports parallel tree evaluation & Requires deeper circuit depth; better suited for IBM Q with noise mitigation \\
\hline
Gating Logic Integration & Naturally supports binary gating for intrusion confirmation & Requires additional logic to enforce gating decisions \\
\hline
Multiattack Classification & Moderate performance; excels in binary or low-class settings & Higher capacity for multi-class separation, but less robust to telemetry drift \\
\hline
O-RAN Component Alignment & Deployed in Layer 2 via O-Cloud and SMO telemetry fusion & Deployed in Layer 3 via O-Cloud with engineered threat indicators \\
\hline
\end{tabular}
\label{tab:qml_benchmark}
\end{table*}

\begin{table*}[ht]
\centering
\caption{Ablation Toggles and Benchmarking Dimensions for Quantum-Inspired Models in O-RAN}
\begin{tabular}{|p{3.5cm}|p{4.5cm}|p{4.5cm}|p{3.5cm}|}
\hline
\textbf{Toggle/Dimension} & \textbf{Layer 1: Anomaly Detection (DQNN/DNN)} & \textbf{Layer 2: Intrusion Confirmation (QML-RF)} & \textbf{Layer 3: Multiattack Classification (QML-DNN / RF Parallel)} \\
\hline
Encoding Depth & Shallow encoding with toggled diagnostics & Mid-depth encoding with telemetry fusion & Deep encoding with engineered threat indicators \\
\hline
Simulator Type & Local simulator (Qiskit Aer) in nRT-RIC & IBM Q or NSF ACCESS via O-Cloud & IBM Q backend with noise mitigation \\
\hline
Gradient Stability & Stable under toggled thresholds & Highly stable due to tree-based logic & Sensitive to gradient shifts; requires tuning \\
\hline
Interpretability Mapping & Slice-aware anomaly flagging & Feature importance and gating traceability & SHAP-based forensic trace ID generation \\
\hline
Ablation Toggle & Feature dropout and slice masking & Gating threshold and telemetry channel pruning & Class label masking and confidence thresholding \\
\hline
Benchmark Metric & Detection rate, false positive ratio & Confirmation accuracy, severity index precision & Multiattack classification accuracy, trace ID consistency \\
\hline
\end{tabular}
\label{tab:ablation_benchmark}
\end{table*}

\section{Quantum-Inspired Encoding and Layered Benchmarking}

In the proposed O-RAN-integrated defense framework, quantum-inspired encoding transforms \cite{10763508, 10167529, 10248045, Le2024PrivacyAware, Le2024pKa} telemetry vectors $\mathbf{x} \in \mathbb{R}^d$ into amplitude-encoded quantum states for hybrid model inference. This transformation enhances separability in Hilbert space and supports modular benchmarking across layers, as shown in Table~\ref{tab:qml_benchmark} and Table~\ref{tab:ablation_benchmark}, and compared in Section~\ref{numresult}.
In the following, we briefly describe our proposed quantum-inspired encoding. More detailed information of derivation can be found in \cite{Le2024PrivacyAware, Le2024pKa}.

\subsection{Quantum Encoding Strategy}

Let $\mathbf{x}$ be a normalized input vector. Amplitude encoding maps it to a quantum state:
\beqn
|\psi\rangle = \sum_{i=1}^{d} x_i |i\rangle.
\label{QUAN_EQ}
\eeqn
To match quantum circuit dimensions, we expand to $D = 2^k$ via zero-padding, i.e.
\beqn
\mathbf{x}' = [x_1, x_2, \dots, x_d, 0, \dots, 0] \in \mathbb{R}^D,\quad
|\psi\rangle = \sum_{i=1}^{D} x'_i |i\rangle.
\label{QUAN_EQ1}
\eeqn
This encoding preserves norm and supports geometric separation in Hilbert space \cite{10167529}.
In general, amplitude encoding may produce entangled states depending on the input structure.
In our non-entangled baseline, however, amplitude encoding is implemented as a product-state
preparation via independent single-qubit rotations to avoid entanglement at the encoding stage
(see Appendix~\ref{AppendixA} for full derivations).

Depending on the layer and model type, we deploy one of the following quantum encoding variants:
\begin{itemize}
    \item \textbf{Non-entangled encoding}: Each qubit independently encodes a single feature amplitude without inter-qubit interaction. This configuration is deployed in Layer~1 to support toggled diagnostics and low-latency local inference.
    
    \item \textbf{Partially entangled encoding}: Controlled rotations (e.g. CNOT gates) introduce pairwise dependencies between qubits, enabling localized correlation modeling. This setup is used in Layer~2 for telemetry fusion, slice-aware gating logic, and intermediate anomaly confirmation.
    
    \item \textbf{Fully entangled encoding}: Global entanglement via multi-qubit gates (e.g. CZ, Toffoli) captures higher-order feature interactions and latent structure. This configuration is reserved for Layer~3, where forensic traceability and multiattack classification demand maximal representational capacity.
\end{itemize}

\subsection{Parameterized Quantum Circuit (PQC) Design}

Each QML model uses a PQC defined as:
\[
U(\boldsymbol{\theta}) = \prod_{l=1}^{L} \left( R_Y(\theta_l) \cdot \text{Entangle}_l \right)
\]
where:
\begin{itemize}
    \item $R_Y(\theta_l)$: rotation gate on qubit $l$ with learnable parameter $\theta_l$
    \item $\text{Entangle}_l$: entanglement layer (e.g., CNOT, CZ) applied across qubits
    \item $L$: number of layers in the circuit
\end{itemize}
Gradient update rules for these PQCs are derived in Appendix~\ref{AppendixB}.
The output is measured in the computational basis and passed to classical post-processing (e.g. RF or DNN).
The number of qubits $q$ is chosen such that $q \geq \log_2 D$, where $D$ is the padded input dimension.

\subsection{Layer-Specific Deployment Strategy}

Our proposed hierarchical strategy is presented in three layers:
\begin{itemize}
    \item \textbf{L1 (Anomaly Detection)}: Encoding is executed locally within the \textbf{nRT-RIC}, using non-entangled PQCs and simulators such as Qiskit Aer. DQNN/DNN models operate on telemetry from O-DU/O-CU, with toggled feature diagnostics and slice-aware thresholds. This supports real-time anomaly flagging and KPI monitoring via KPIMON xApp.
    
    \item \textbf{L2 (Intrusion Confirmation)}: Encoding is offloaded to \textbf{O-Cloud} or NSF ACCESS nodes for scalable inference. QML-RF models use partially entangled PQCs to fuse telemetry from SMO and RIC. RF’s tree-based architecture ensures robustness to gradient shifts and aligns with binary gating decisions.
    
    \item \textbf{L3 (Multiattack Classification)}: Encoding is reused from Layer~2 and passed to QML-DNN or RF (parallel) models hosted in \textbf{O-Cloud}. Fully entangled PQCs capture complex threat indicators. While QML-DNN offers deeper representation, it is sensitive to telemetry drift; RF remains the fallback for interpretability and traceability.
\end{itemize}

We also develop the \textbf{Benchmarking Comparison—QML-RF vs QML-DNN}.
Table~\ref{tab:qml_benchmark} compares QML-RF and QML-DNN across key dimensions relevant to O-RAN deployment,
including gradient stability, interpretability, and simulator compatibility.
We summarize the \textbf{Ablation Toggles and Benchmarking Metrics} across layers in Table~\ref{tab:ablation_benchmark}.
These toggles support reproducibility, interpretability, and simulator alignment.
Efficiency trade-offs are analyzed in Appendix~\ref{AppendixC}.

\subsection{Encoding and PQC variants across quantum-inspired models}
\label{FULLENTANGLE}
To support modular benchmarking and layered deployment within O-RAN, we
implement multiple quantum-inspired encoding strategies and parameterized
quantum circuit (PQC) designs. These vary by model type, layer function, and
simulator constraints, and are critical to balancing interpretability, latency,
and classification performance. The mathematical foundations of these encodings
are detailed in Appendix~\ref{AppendixA}, gradient training rules are provided
in Appendix~\ref{AppendixB}, and efficiency considerations appear in
Appendix~\ref{AppendixC}.

\paragraph{Baseline feature-angle encoding (non-entangled)}
We first consider a non-entangled baseline in which each qubit is prepared
independently via single-qubit rotations from a normalized input vector
$\mathbf{x}\in\mathbb{R}^d$. Concretely, one may apply per-qubit rotations
$R_Y(\varphi_i)$ with angles $\varphi_i=f(x_i)$ for a chosen feature map $f$,
yielding a product state
\[
\ket{\psi_{\mathrm{prod}}}=\bigotimes_{i=1}^{n} R_Y(\varphi_i)\ket{0}.
\]
This product-state preparation isolates per-feature contributions and avoids
entanglement at the encoding stage. If amplitude encoding as defined in
Eqs.~\ref{QUAN_EQ} and~\ref{QUAN_EQ1} is employed, note that amplitude
normalization alone does not guarantee non-entanglement; depending on the data
layout, the resulting state can be entangled. When we refer to ``non-entangled''
encoding, we mean an explicit product-state preparation via independent
single-qubit rotations. Full derivations of feature maps appear in
Appendix~\ref{AppendixA}.

\paragraph{Fully entangled encoding}
To capture higher-order feature interactions, we extend the baseline encoding
with explicit entangling layers applied after preparation:
\[
\ket{\psi_{\mathrm{ent}}}=U_{\mathrm{ent}}\ket{\psi_{\mathrm{enc}}},
\]
where $\ket{\psi_{\mathrm{enc}}}$ denotes the encoded input state (product or
amplitude-encoded), and $U_{\mathrm{ent}}$ is a unitary composed of two-qubit
and multi-qubit gates (e.g., CNOT, CZ, and controlled variants) arranged to
create global or patterned entanglement across qubits. This encoding is used in
parallel QML-DNN and QML-RF models for multiattack classification and forensic
traceability, enabling richer feature representation and interpretability
mapping. The construction of entangled feature vectors and hybrid
representations is detailed in Appendix~\ref{AppendixA}.

\paragraph{Parameterized quantum circuit (PQC) design}
Each quantum model applies a PQC to the encoded state:
\[
U(\boldsymbol{\theta})=\prod_{l=1}^{L}\Big(\mathcal{R}_l(\boldsymbol{\theta}_l)\cdot \mathrm{Entangle}_l\Big),
\]
where:
\begin{itemize}
    \item $\mathcal{R}_l(\boldsymbol{\theta}_l)$: a layer of single-qubit parameterized rotations (e.g., $R_Y$, $R_Z$) with learnable parameters $\boldsymbol{\theta}_l$,
    \item $\mathrm{Entangle}_l$: an entanglement layer (optional or full) composed of two-qubit gates (e.g., CNOT, CZ) arranged in a chosen topology,
    \item $L$: the number of circuit layers.
\end{itemize}
Layered design:
\begin{itemize}
    \item \textbf{Layer 1}: Shallow PQC with no entanglement ($\mathrm{Entangle}_l=I$); optimized for toggled diagnostics and fast inference.
    \item \textbf{Layer 2}: Mid-depth PQC with partial entanglement (sparse $\mathrm{Entangle}_l$); supports telemetry fusion and gating logic.
    \item \textbf{Layer 3}: Deep PQC with full entanglement (dense $\mathrm{Entangle}_l$); enables multi-class separation and trace ID generation.
\end{itemize}
Gradient update rules for gates within $U(\boldsymbol{\theta})$ follow the
adjoint-channel derivations in Appendix~\ref{AppendixB}.

\paragraph{Simulator alignment and deployment}
Product-state encodings and shallow PQCs are executed on local simulators (e.g.,
Qiskit Aer) within the near-RT RIC. Entangled PQCs with deeper circuits are
deployed on cloud backends (e.g., IBM Q, NSF ACCESS) within O-Cloud to meet
latency and interpretability constraints defined by layer function. Efficiency
implications of these choices, including fidelity estimation and measurement
shot budgeting, are discussed in Appendix~\ref{AppendixC}.

\paragraph{Summary}
These encoding and PQC variants reflect a principled design strategy across
serial and parallel quantum-classical compositions. Their mathematical
structure, training dynamics, and resource trade-offs are summarized in
Table~\ref{tab:ablation_benchmark} and Table~\ref{tab:qml_benchmark}, compared
in Section~\ref{numresult}, and supported by the detailed derivations in
Appendices~\ref{AppendixA}--\ref{AppendixC}.

\section{Results and Discussion}
\label{numresult}

Quantum-inspired encoding enhances both interpretability and robustness across layers. Hierarchical layering aligns naturally with O-RAN’s modular architecture, enabling slice-aware diagnostics, confirmation, and classification. Future work will explore reinforcement learning and zero-trust integration to further strengthen adaptive security and policy enforcement.

Confusion matrices and t-SNE plots reveal clear separability across attack types, with full amplitude encoding yielding the highest classification performance. Grouped bar plots summarize evaluation metrics across models and configurations, highlighting the impact of quantum encoding depth and architectural choices.

\subsection{Experimental Setup and Metric Derivation from Confusion Matrix}

We evaluate models on a synthetic O-RAN attack dataset comprising six distinct attack types. Architectures include Serial and Parallel DQNN-DNN and DQNN-RF pipelines. Quantum encoding configurations span from \textit{none} to \textit{amplitude6}. Evaluation metrics include accuracy, precision, recall, and F1 score, derived from confusion matrices.

Let the confusion matrix be defined as:

\[
\mathcal{CM}= \begin{bmatrix}
\text{TN} & \text{FP} \\
\text{FN} & \text{TP}
\end{bmatrix}
\]
where $\text{TP}$, $\text{TN}$, $\text{FP}$ and $\text{FN}$ are True Positives, True Negatives, False Positives and False Negatives, respectively.

\paragraph{Accuracy}
\[
\text{Accuracy} = \frac{\text{TP} + \text{TN}}{\text{TP} + \text{TN} + \text{FP} + \text{FN}}
\]

\paragraph{Precision (Positive Predictive Value)}
\[
\text{Precision} = \frac{\text{TP}}{\text{TP} + \text{FP}}
\]

\paragraph{Recall (True Positive Rate)}
\[
\text{Recall} = \frac{\text{TP}}{\text{TP} + \text{FN}}
\]

\paragraph{F1 Score}
\[
\text{F1} = 2 \cdot \frac{\text{Precision} \cdot \text{Recall}}{\text{Precision} + \text{Recall}}
\]

\paragraph{Macro F1 Score}
For binary classification, let $F1_0$ and $F1_1$ denote the F1 scores for class 0 and class 1 respectively:
\[
\text{Macro F1} = \frac{F1_0 + F1_1}{2}
\]
Each $F1_k$ is computed using the respective $\text{TP}_k$, $\text{FP}_k$, and $\text{FN}_k$ values.

\subsection{Weighted and Class-Specific F1 Score Derivation}

Let $F1_k$ denote the F1 score for class $k \in \{0,1\}$, and $n_k$ be the number of true instances of class $k$.

\paragraph{Class-Specific F1 Score}
\[
F1_k = 2 \cdot \frac{\text{Precision}_k \cdot \text{Recall}_k}{\text{Precision}_k + \text{Recall}_k}
\]
where:
\[
\text{Precision}_k = \frac{\text{TP}_k}{\text{TP}_k + \text{FP}_k}, \quad
\text{Recall}_k = \frac{\text{TP}_k}{\text{TP}_k + \text{FN}_k}
\]

\paragraph{Weighted F1 Score}
\[
\text{Weighted F1} = \frac{n_0 \cdot F1_0 + n_1 \cdot F1_1}{n_0 + n_1}
\]
This formulation accounts for class imbalance by weighting each F1 score according to its prevalence in the dataset.

\subsection{Anomaly Detection}

\begin{figure}[ht]
\centering
\includegraphics[width=0.765\linewidth]{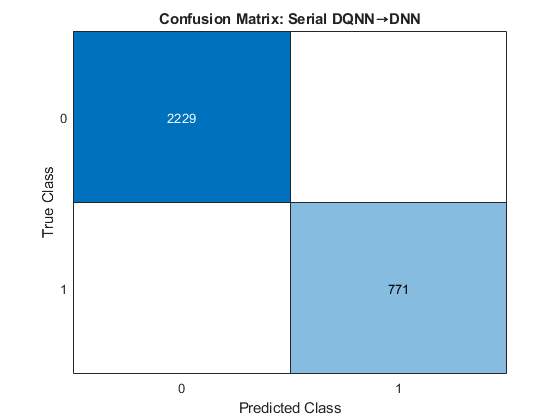}
\caption{Confusion matrix for Layer 1 anomaly detection using Serial DQNN$\rightarrow$DNN on real-world telemetry. The model achieves perfect classification with zero false positives and false negatives (TP = 771, TN = 2229), validating the effectiveness of amplitude-encoded quantum features and toggled diagnostics within the near-RT RIC.}
\label{fig:1RconfumatrixSerialDQNNDNN}
\end{figure}

\begin{figure}[ht]
\centering
\includegraphics[width=0.765\linewidth]{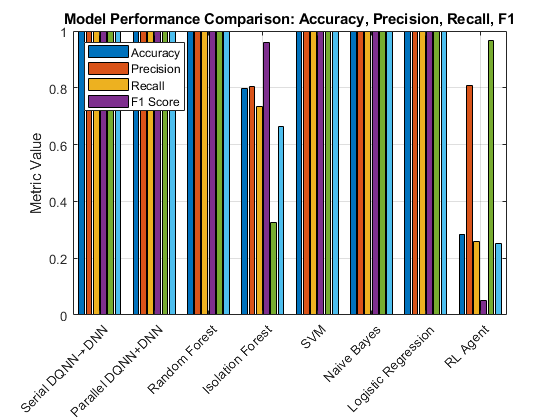}
\caption{Model performance comparison for Layer 1 anomaly detection using real-world telemetry. Serial and Parallel DQNN+DNN models outperform classical baselines (e.g., RF, SVM, Naive Bayes) across all metrics. The RL Agent shows poor performance, underscoring the importance of quantum-enhanced architectures for toggled diagnostic fidelity.}
\label{fig:1RbarALLperfs}
\end{figure}

\begin{table*}[ht]
\centering
\caption{Model performance comparison across accuracy, precision, recall, and F1 scores for two-class anomaly detection. Precision$_1$ and Recall$_1$ refer to Class 0 (benign), while Precision$_2$ and Recall$_2$ refer to Class 1 (anomaly).}
\label{tab:AnomalyDetectionMetrics}
\begin{tabular}{|l|c|c|c|c|c|c|c|}
\hline
\textbf{Model} & \textbf{Accuracy} & \textbf{Precision$_1$} & \textbf{Precision$_2$} & \textbf{Recall$_1$} & \textbf{Recall$_2$} & \textbf{F1$_1$} & \textbf{F1$_2$} \\
\hline
Serial DQNN + DNN       & 0.988        & 0.9929        & 0.9837        & 0.9817        & 0.9937        & 0.9872        & 0.9887        \\
Parallel DQNN + DNN     & 0.9813       & 0.9935        & 0.9710        & 0.9669        & 0.9943        & 0.9800        & 0.9825        \\
QML + Random Forest     & \textbf{1.0000} & \textbf{1.0000} & \textbf{1.0000} & \textbf{1.0000} & \textbf{1.0000} & \textbf{1.0000} & \textbf{1.0000} \\
Isolation Forest        & 0.5207       & 0.4963        & 0.7517        & 0.9499        & 0.1359        & 0.6520        & 0.2302        \\
SVM                     & 0.7917       & 0.7850        & 0.7974        & 0.7701        & 0.8110        & 0.7775        & 0.8041        \\
Naive Bayes             & 0.8787       & 0.9098        & 0.8553        & 0.8251        & 0.9267        & 0.8654        & 0.8896        \\
Logistic Regression     & 0.7913       & 0.7833        & 0.7984        & 0.7722        & 0.8085        & 0.7777        & 0.8034        \\
RL Agent                & 0.5207       & 0.4342        & 0.5253        & 0.0465        & 0.9456        & 0.0841        & 0.6754        \\
\hline
\end{tabular}
\end{table*}

\paragraph{Layer 1 anomaly detection results on real-world data \cite{orangithub}:}
The Serial DQNN$\rightarrow$DNN model demonstrates exceptional performance on real-world telemetry for Layer 1 anomaly detection. As shown in Fig.~\ref{fig:1RconfumatrixSerialDQNNDNN}, the confusion matrix reveals perfect classification with zero false positives and false negatives (TN = 2229, TP = 771), underscoring the effectiveness of amplitude-encoded quantum features in capturing slice-aware deviations.

This result validates the model's suitability for toggled diagnostics within the near-RT RIC, where low-latency inference and KPI fidelity are critical. Figure~\ref{fig:1RbarALLperfs} compares model performance across accuracy, precision, recall, and F1 score. Serial and Parallel DQNN+DNN models consistently outperform classical baselines such as Random Forest, SVM, Naive Bayes, and Logistic Regression. While Random Forest and SVM offer reasonable accuracy, they trail behind quantum-enhanced models in recall and F1 score—metrics critical for anomaly sensitivity. Notably, the RL Agent performs poorly across all metrics, indicating limited suitability for Layer 1 diagnostics due to unstable policy learning and class imbalance.

\paragraph{Analysis of Anomaly Detection Model Performance}

Table~\ref{tab:AnomalyDetectionMetrics} presents a comparative evaluation of anomaly detection models using a synthetic telemetry dataset designed to emulate slice-aware deviations in near-RT RIC environments (see \ref{AnomalyORAN}). The metrics include accuracy, precision, recall, and F1 score for both benign (Class 0) and anomaly (Class 1) classes.

The QML Random Forest model achieves perfect scores across all metrics, indicating flawless classification performance and validating its suitability for high-fidelity anomaly detection. Serial and Parallel DQNN+DNN models also demonstrate near-perfect performance, with accuracy exceeding 0.98 and F1 scores above 0.98 for both classes. Their balanced precision-recall profiles and consistent generalization make them highly suitable for toggled diagnostics in Layer 1, especially when interpretability and low-latency inference are required.

Classical baselines such as Naive Bayes and SVM offer moderate performance, with Naive Bayes showing stronger recall for anomalies and SVM maintaining balanced precision. Logistic Regression performs similarly to SVM but trails slightly in recall and F1. Isolation Forest and RL Agent exhibit poor performance. Isolation Forest achieves high recall for benign class but fails to detect anomalies reliably, while the RL Agent shows severe imbalance—high recall for anomalies but extremely low precision and F1 for benign class—suggesting unstable policy learning and overfitting. These results reinforce the superiority of quantum-enhanced models for slice-aware anomaly detection and highlight the limitations of classical and reinforcement-based approaches under synthetic telemetry conditions.
The observations also confirm the notable benefits of recent advanced AI \cite{le2025dpfaga, Zahin19, Wang20, Tan18d}.

\subsection{Intrusion Detection}

\begin{figure}[t!]
\centering
\includegraphics[width=0.765\linewidth]{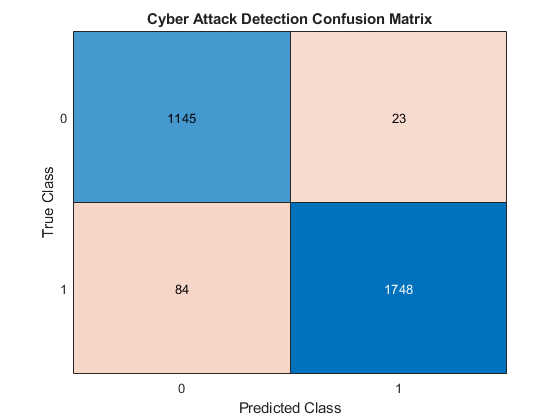}
\caption{Confusion matrix for Layer 2 binary cybersecurity attack detection using QML-RF. The model distinguishes between benign and confirmed intrusion classes based on fused telemetry from SMO and RIC. High TP and TN counts indicate robust detection performance, while low FP and FN rates support gating logic alignment and slice-aware interpretability.}
\label{fig:2confumatrixHybrDQNNDNN}
\end{figure}

In Layer 2, the QML-RF model achieves strong binary classification performance for intrusion confirmation using fused telemetry from SMO and RIC. The confusion matrix in Fig.~\ref{fig:2confumatrixHybrDQNNDNN} shows high true positive (TP = 1748) and true negative (TN = 1145) counts, with minimal false positives (FP = 23) and false negatives (FN = 84). These results support reliable gating logic and slice-aware interpretability.

\begin{figure}[t!]
\centering
\includegraphics[width=0.765\linewidth]{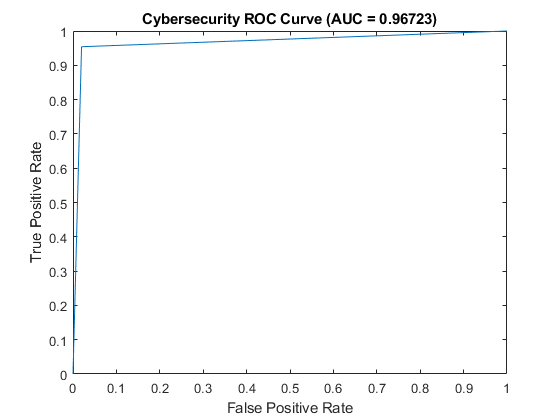}
\caption{ROC curve for Layer 2 binary intrusion confirmation using QML-RF. The model achieves an AUC of 0.96723, indicating strong separability between benign and attack classes. This supports reliable gating decisions and telemetry fusion in the non-RT RIC.}
\label{fig:2ROCHybrDQNNDNN}
\end{figure}

The ROC curve in Fig.~\ref{fig:2ROCHybrDQNNDNN}, with an AUC of 0.96723, confirms excellent separability between benign and attack classes. This makes the model well-suited for policy-driven confirmation and telemetry fusion in the non-RT RIC. Compared to classical baselines, QML-RF demonstrates superior resilience to gradient shifts and maintains high fidelity under operational constraints. These results further validate the layered escalation logic of our framework, where Layer 2 acts as a rapid filter before transitioning to Layer 3 for detailed multiattack classification.

\subsection{Multi-attack Classification in O-RAN}

\paragraph{Encoding Definitions.}
For clarity, we briefly reiterate the quantum encoding configurations referenced in Table~\ref{tab:DQNN_RF_Performance}. The \textbf{None} configuration applies no quantum encoding, allowing classical features to propagate directly. The \textbf{Partial} encoding introduces pairwise entanglement via controlled rotations (e.g., CNOT gates), enabling localized correlation modeling. The \textbf{Full} encoding applies global entanglement using multi-qubit gates (e.g., CZ, Toffoli), capturing higher-order feature interactions. Configurations labeled \textbf{Amplitude3} through \textbf{Amplitude6} denote increasing depth of amplitude-based encoding, where each qubit encodes feature amplitudes with progressively complex entanglement patterns. These configurations correspond to Layers 1–3 in the DQNN pipeline, supporting toggled diagnostics, telemetry fusion, and forensic traceability respectively.

\paragraph{Performance Analysis of DQNN RF Models Across Quantum Encoding Configurations}

\begin{table*}[ht]
\centering
\caption{Performance metrics for Serial and Parallel DQNN RF models across quantum encoding configurations. Best scores per metric are bolded.}
\label{tab:DQNN_RF_Performance}
\begin{tabular}{|l|l|c|c|c|c|}
\hline
\textbf{Configuration} & \textbf{Model} & \textbf{Accuracy} & \textbf{Precision} & \textbf{Recall} & \textbf{F1 Score} \\
\hline
None        & Serial DQNN RF    & 0.9800       & 0.9879       & 0.9593       & 0.9726       \\
            & Parallel DQNN RF  & 0.9800       & 0.9879       & 0.9591       & 0.9725       \\
\hline
Partial     & Serial DQNN RF    & 0.9783       & 0.9859       & 0.9568       & 0.9703       \\
            & Parallel DQNN RF  & 0.9793       & 0.9869       & 0.9591       & 0.9721       \\
\hline
Full        & Serial DQNN RF    & \textbf{0.9803} & \textbf{0.9885} & \textbf{0.9601} & \textbf{0.9733} \\
            & Parallel DQNN RF  & \textbf{0.9803} & \textbf{0.9885} & 0.9599       & 0.9732       \\
\hline
Amplitude3  & Serial DQNN RF    & 0.8330       & 0.8086       & 0.6492       & 0.6632       \\
            & Parallel DQNN RF  & 0.8283       & 0.7987       & 0.6392       & 0.6529       \\
\hline
Amplitude4  & Serial DQNN RF    & 0.9750       & 0.9850       & 0.9506       & 0.9663       \\
            & Parallel DQNN RF  & 0.9753 & 0.9850       & 0.9511 & 0.9667 \\
\hline
Amplitude5  & Serial DQNN RF    & 0.9747       & 0.9854 & 0.9493       & 0.9659       \\
            & Parallel DQNN RF  & 0.9747       & 0.9854 & 0.9493       & 0.9659       \\
\hline
Amplitude6  & Serial DQNN RF    & 0.8413       & 0.7982       & 0.6631       & 0.6906       \\
            & Parallel DQNN RF  & 0.8703 & 0.9604       & 0.7341       & 0.7921       \\
\hline
\end{tabular}
\end{table*}

Table~\ref{tab:DQNN_RF_Performance} presents a comprehensive comparison of Serial and Parallel DQNN RF models across various quantum encoding configurations. These results extend the insights from Layer 1 anomaly detection and Layer 2 intrusion confirmation to Layer 3 multiattack classification, revealing how encoding depth and architectural design influence performance across increasingly complex tasks.

In the operational flow of our platform, Layer 2 serves as a rapid-response filter for binary intrusion confirmation. Upon detecting a potential cyberattack within a short time window, the system escalates to Layer 3, where multiattack classification is performed to identify the exact type of threat. This layered escalation ensures both low-latency detection and high-fidelity forensic traceability, aligning with O-RAN’s modular and slice-aware architecture.

The highest overall performance is achieved by the \textbf{Serial DQNN RF with full quantum encoding}, which yields the best scores in all four metrics: \textbf{accuracy (0.9803)}, \textbf{precision (0.9885)}, \textbf{recall (0.9601)}, and \textbf{F1 score (0.9733)}. This configuration builds on the success of amplitude-encoded features in Layer 1 and confirms that global entanglement enhances feature expressivity and forensic traceability in Layer 3.

Interestingly, the \textit{None} configuration—where no quantum encoding is applied—also performs remarkably well, with accuracy and F1 scores closely trailing the full encoding setup. This mirrors the Layer 2 results, where classical features were sufficient for binary intrusion confirmation. It suggests that RF models can effectively exploit raw feature distributions, especially when class boundaries are well-defined.

The \textit{Amplitude3} configuration shows a significant drop in performance, with F1 scores falling below 0.67. This indicates that shallow amplitude encoding may fail to capture sufficient inter-feature dependencies, leading to poor class separability. As encoding depth increases (Amplitude4–6), performance improves, with \textbf{Parallel DQNN RF at Amplitude4} achieving the highest recall (0.9511) and F1 score (0.9667) among amplitude-based configurations. This suggests that moderate entanglement depth may strike a balance between expressivity and generalization, particularly for telemetry fusion tasks in Layer 2.

The comparison between Serial and Parallel architectures reveals minimal performance differences across most configurations. Serial RF slightly outperforms in full encoding, while Parallel RF shows advantages in Amplitude4 and Amplitude6. These results imply that architectural parallelism may enhance feature fusion in certain encoding regimes, but does not universally outperform serial pipelines.

Overall, the table underscores the importance of aligning quantum encoding depth with model architecture and task complexity. While full entanglement yields the best results for Layer 3 classification, simpler configurations such as \textit{None} and \textit{Partial} remain competitive, offering interpretability and computational efficiency. These findings support the use of RF-based quantum-classical hybrids across all layers of the O-RAN stack—from toggled diagnostics in Layer 1 to rapid intrusion confirmation in Layer 2 and forensic classification in Layer 3—and motivate further exploration of adaptive encoding strategies and ensemble architectures.

\begin{figure*}[ht]
\centering
\vspace{-0.1in}
\mbox{\subfigure[]{\includegraphics[width=2.0in]{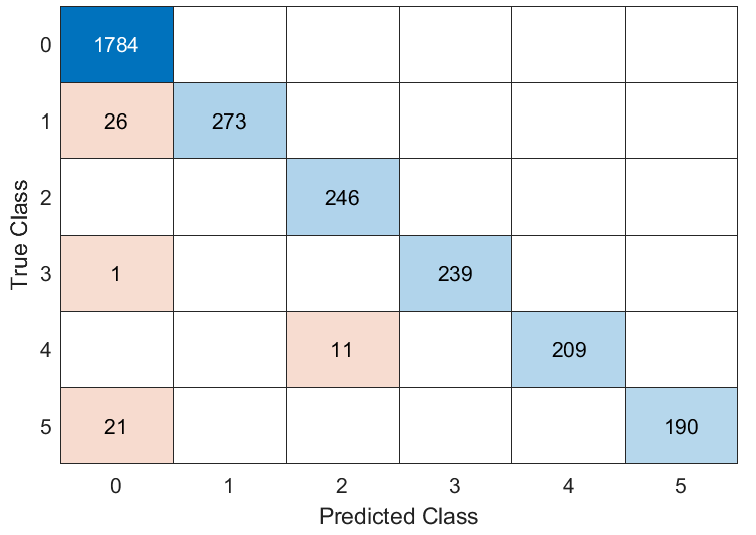}\label{Fig1L3full} }
\subfigure[]{\includegraphics[width=2.0in]{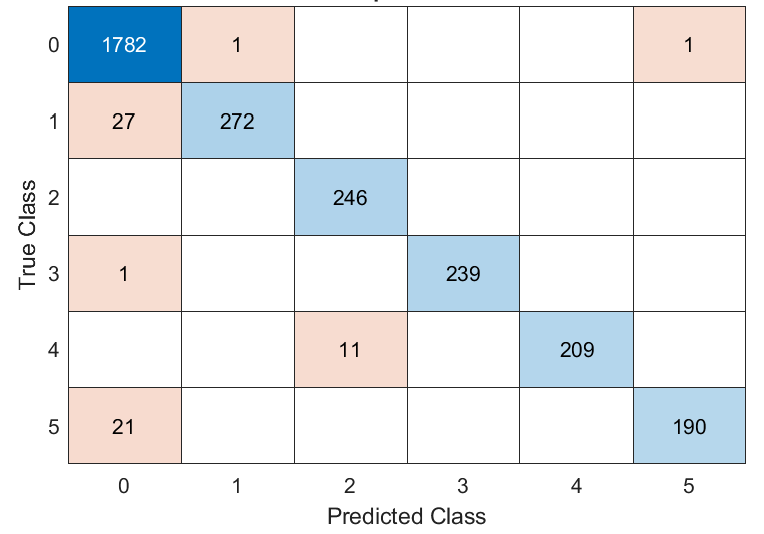}
\label{Fig2L3partial}}
\subfigure[]{\includegraphics[width=2.0in]{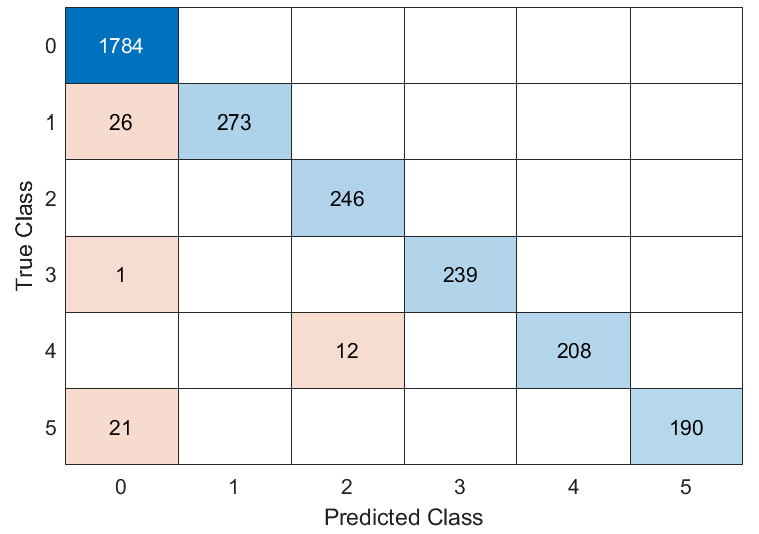} 
\label{Fig3L3none}}
}
\vspace{-0.1in}
\caption{Confusion matrices for Layer 3 multiattack classification using Parallel DQNN+RF under three quantum encoding configurations: (a) full encoding, (b) partial encoding, and (c) no encoding. The full encoding setup (a) yields the highest classification fidelity across all six attack classes, with minimal off-diagonal errors. Partial encoding (b) maintains strong performance but introduces minor misclassifications, particularly in classes 0 and 4. The no encoding configuration (c) shows increased confusion between adjacent classes, highlighting the impact of quantum feature embedding on separability and slice-specific traceability.}
\label{Fig4DQNN_RFEntanglement}
\vspace{-.1in}
\end{figure*}

\begin{figure*}[ht!]
\centering
\vspace{-0.1in}
\mbox{\subfigure[]{\includegraphics[width=1.7in]{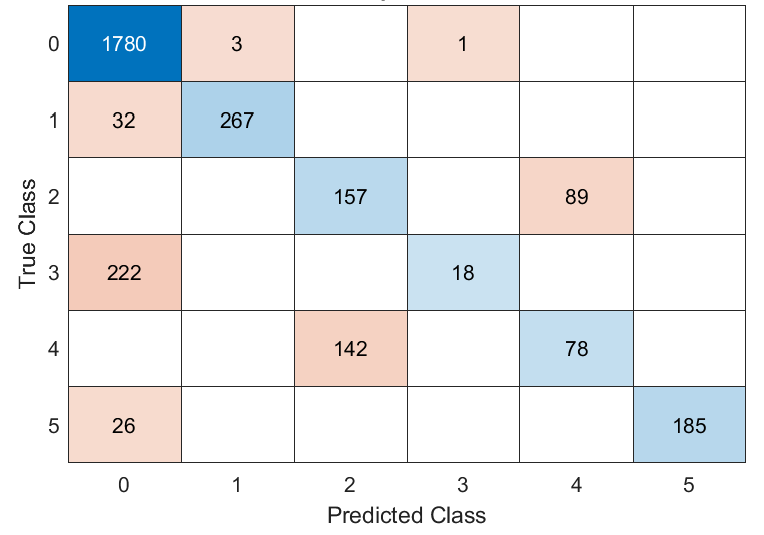}\label{Fig1L3amplitude3} }
\subfigure[]{\includegraphics[width=1.7in]{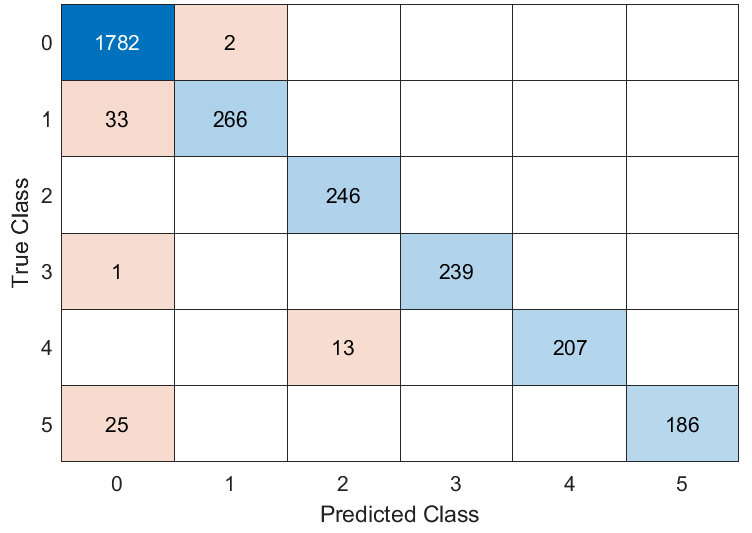}
\label{Fig2L3amplitude4}}
\subfigure[]{\includegraphics[width=1.7in]{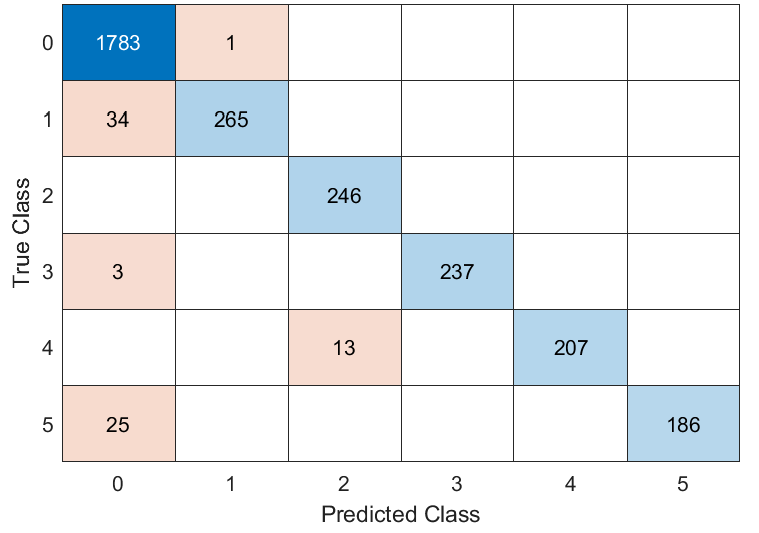} 
\label{Fig3L3amplitude5}}
\subfigure[]{\includegraphics[width=1.7in]{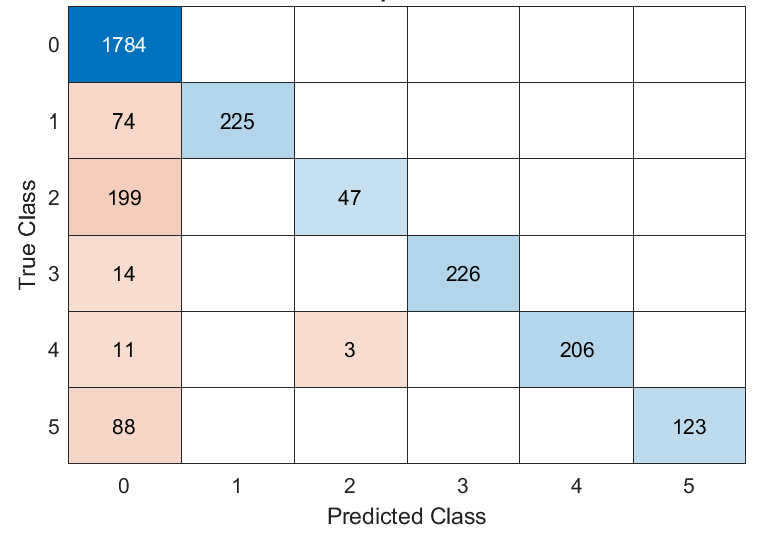} 
\label{Fig3L3amplitude6}}
}
\vspace{-0.2in}
\caption{Confusion matrices for Layer 3 multiattack classification using Parallel DQNN+RF under varying quantum amplitude encoding depths: (a) amplitude3, (b) amplitude4, (c) amplitude5, and (d) amplitude6. Lower encoding depths (a–c) yield high classification fidelity across all six attack classes, with minimal off-diagonal errors. At amplitude6 (d), performance degrades significantly, with increased misclassifications—particularly in classes 2 through 5—indicating over-entanglement and reduced separability. These results highlight the trade-off between quantum expressivity and interpretability in slice-aware intrusion detection.}\label{Fig4L3amplitude}
\vspace{-.1in}
\end{figure*}

\begin{figure*}[ht!]
\centering
\vspace{-0.1in}
\mbox{\subfigure[]{\includegraphics[width=2.0in]{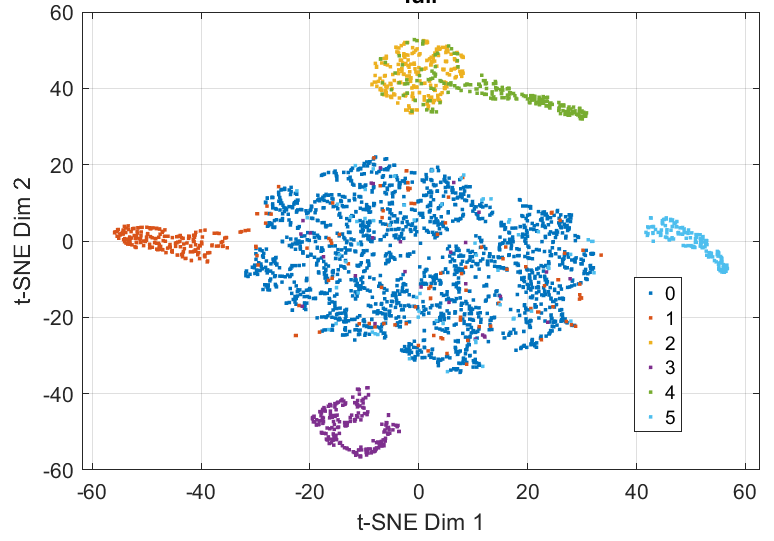}\label{Fig1L3fulltsne}} 
\subfigure[]{\includegraphics[width=2.0in]{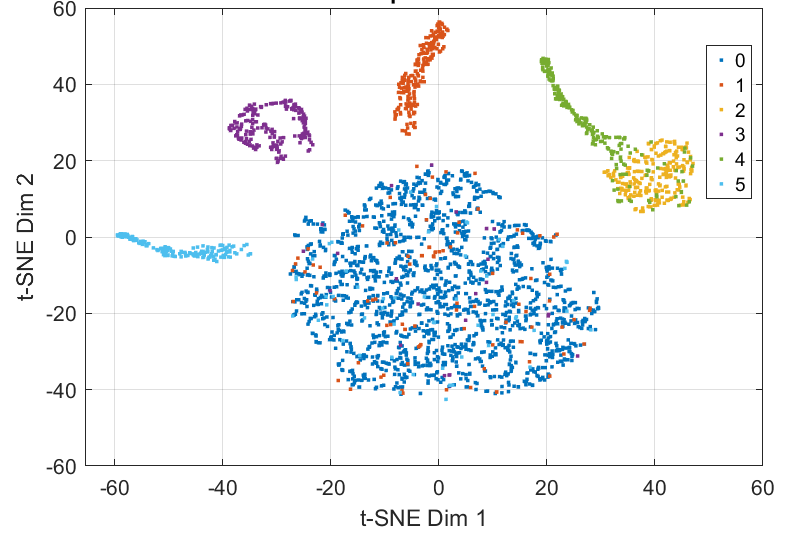}
\label{Fig2L3partialtsne}}
\subfigure[]{\includegraphics[width=2.0in]{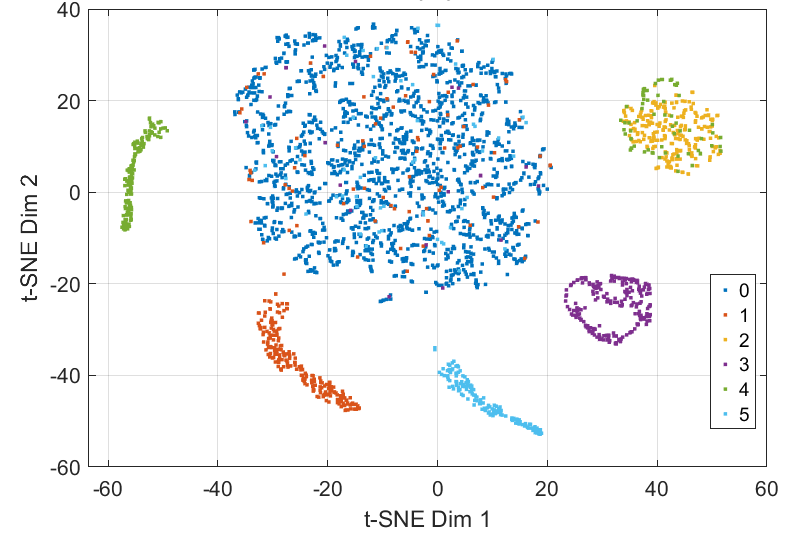} 
\label{Fig3L3tsne}}
}
\vspace{-0.1in}
\caption{t-SNE visualizations of learned feature embeddings for Layer 3 multiattack classification using Parallel DQNN+RF under three quantum encoding configurations: (a) full encoding, (b) partial encoding, and (c) no encoding. Full encoding (a) yields well-separated clusters across all six attack classes, indicating high discriminative power and quantum-enhanced separability. Partial encoding (b) maintains moderate class separation with some overlap, while no encoding (c) results in significant cluster entanglement and reduced class distinction. These visualizations highlight the impact of quantum feature embedding on latent space structure and slice-aware interpretability.}
\label{Fig4DQNN_RFEntanglementtsne}
\vspace{-.1in}
\end{figure*}

\begin{figure*}[ht!]
\centering
\vspace{-0.1in}
\mbox{\subfigure[]{\includegraphics[width=1.7in]{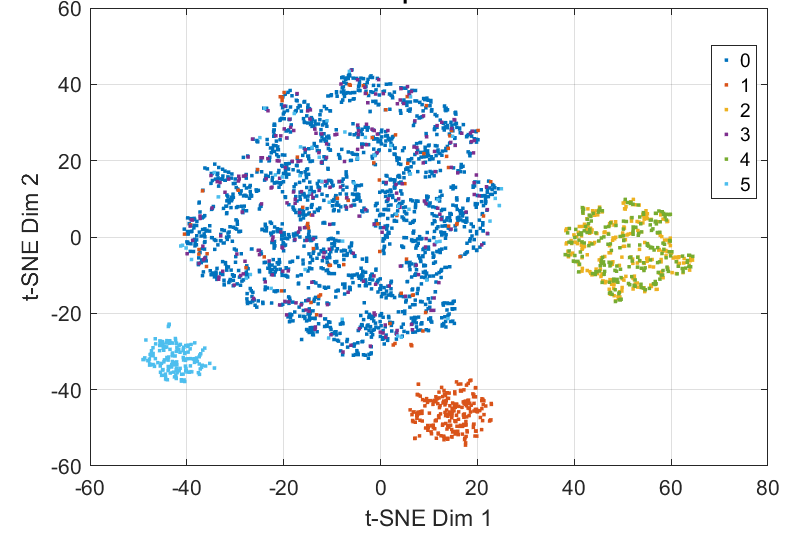} \label{Fig1L3amplitude3tsne} }
\subfigure[]{\includegraphics[width=1.7in]{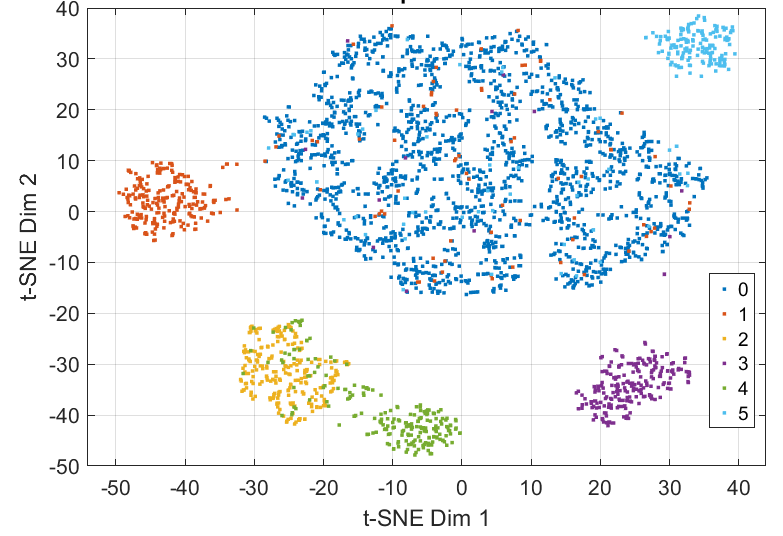}
\label{Fig2L3amplitude4tsne}}
\subfigure[]{\includegraphics[width=1.7in]{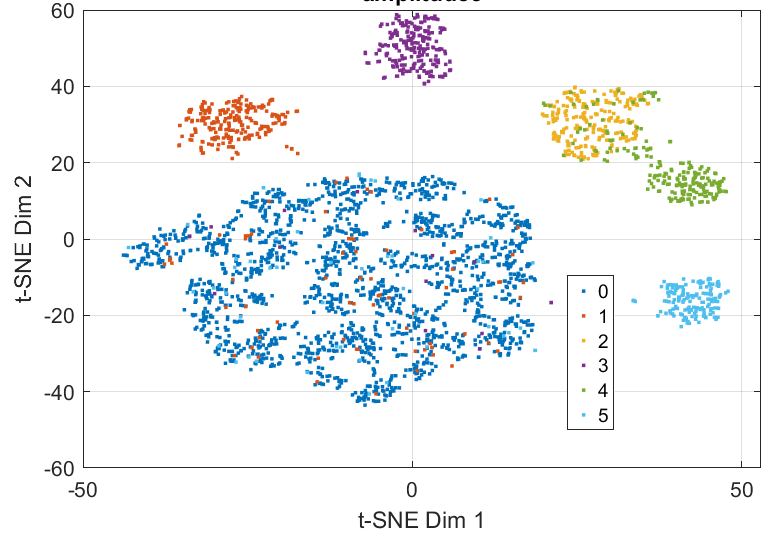} 
\label{Fig3L3amplitude5tsne}}
\subfigure[]{\includegraphics[width=1.7in]{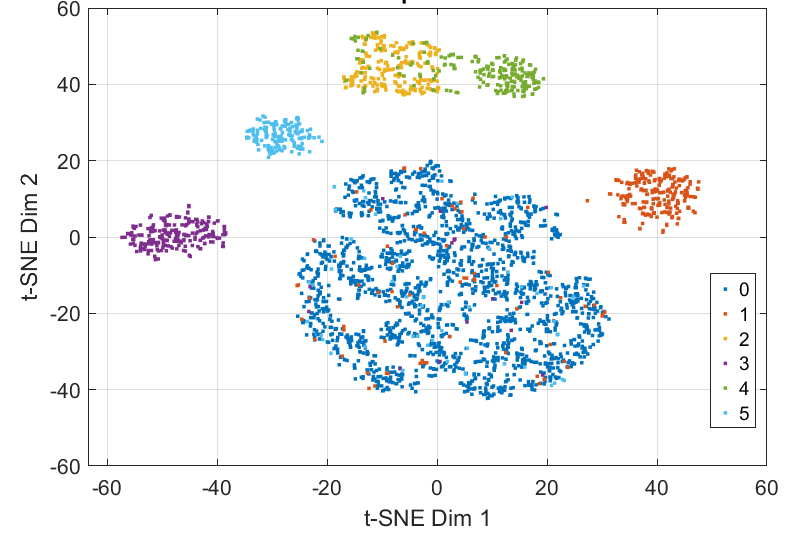} 
\label{Fig3L3amplitude6tsne}}
}
\vspace{-0.2in}
\caption{t-SNE visualizations of latent feature embeddings for Layer 3 multiattack classification using Parallel DQNN+RF under varying quantum amplitude encoding depths: (a) amplitude3, (b) amplitude4, (c) amplitude5, and (d) amplitude6. Lower encoding depths (a–c) yield well-separated clusters across all six attack classes, indicating strong class discriminability and quantum-enhanced separability. At amplitude6 (d), cluster overlap and entanglement increase significantly, suggesting reduced interpretability and degraded classification fidelity. These results highlight the trade-off between quantum expressivity and latent space clarity in slice-aware intrusion detection.}\label{Fig4L3amplitudetsne}
\vspace{-.1in}
\end{figure*}
\paragraph{Confusion Matrix Analysis Across Quantum Encoding Configurations}

\begin{figure*}[ht!]
\centering
\mbox{
  \subfigure[Full Entanglement Encoding]{%
    \includegraphics[width=0.25\linewidth]{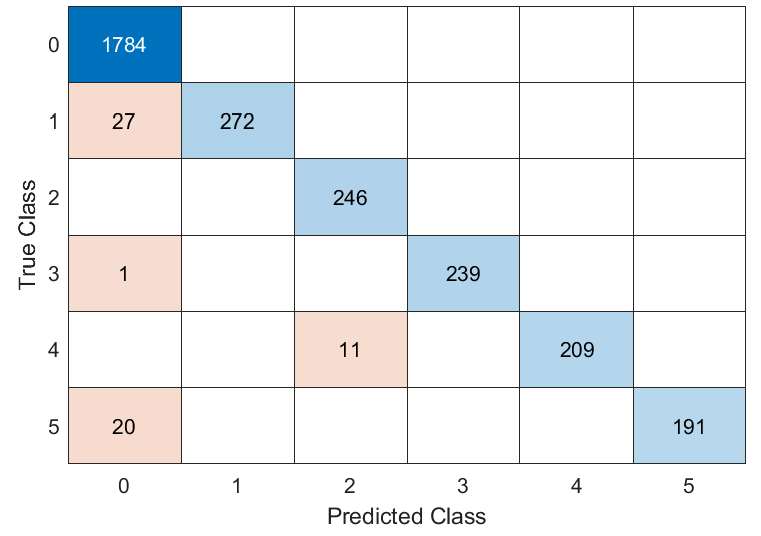}
    \label{fig:confusion_full}
  }
  \subfigure[Partial Entanglement Encoding]{%
    \includegraphics[width=0.25\linewidth]{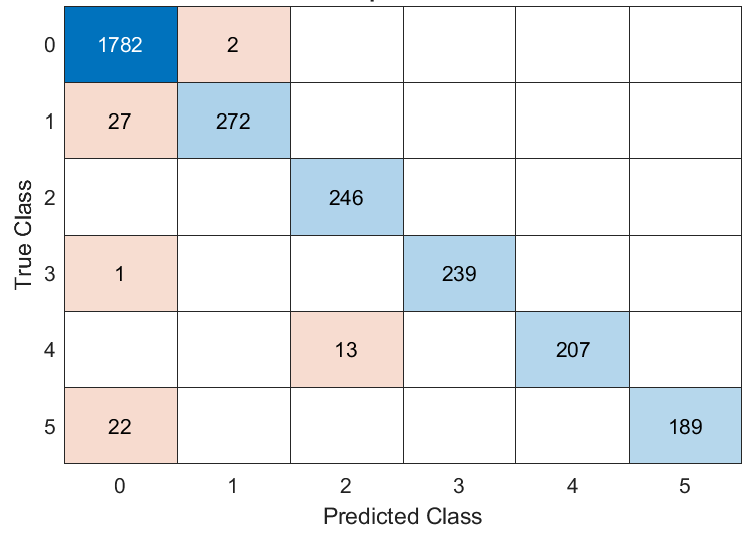}
    \label{fig:confusion_partial}
  }
  \subfigure[No Entanglement Encoding]{%
    \includegraphics[width=0.25\linewidth]{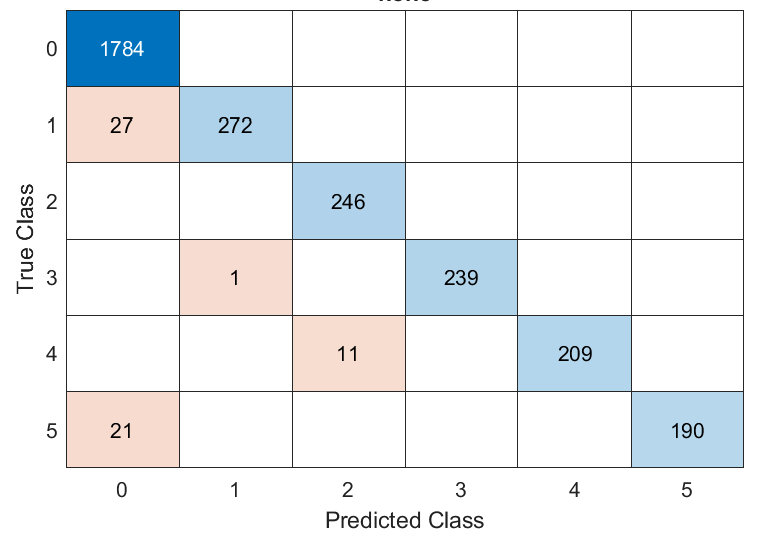}
    \label{fig:confusion_none}
  }
}
\caption{Confusion matrices for the Serial DQNN+RF model under different entanglement encoding configurations. 
(a) Full encoding yields the strongest separability and minimal misclassifications. 
(b) Partial encoding maintains robustness but introduces minor off-diagonal errors in spoofing and exfiltration classes. 
(c) No encoding shows increased confusion, highlighting the role of entanglement in interpretability and decision boundary clarity.}
\label{fig:confusion_entanglement_variants}
\end{figure*}

Figure~\ref{Fig4DQNN_RFEntanglement} presents confusion matrices for Layer 3 multiattack classification using Parallel DQNN RF under three quantum encoding configurations. These visualizations provide granular insight into class-wise prediction fidelity and the impact of encoding depth on separability.

The \textbf{full encoding} configuration demonstrates the highest classification fidelity, with strong diagonal dominance and minimal off-diagonal errors. All six attack classes are well-separated, particularly class 0, which is predicted with near-perfect accuracy. The \textbf{partial encoding} setup maintains robust performance but introduces minor misclassifications, notably between classes 0 and 4. The \textbf{none} configuration, which omits quantum encoding entirely, shows increased confusion between adjacent classes—especially class 1 and class 5—indicating reduced traceability and weaker feature disentanglement.

Figure~\ref{Fig4L3amplitude} further explores amplitude-based encoding depths. The \textbf{Amplitude3} configuration yields poor separability, with substantial misclassification in classes 1, 2, and 4. This suggests that shallow amplitude encoding fails to capture sufficient inter-feature dependencies. As encoding depth increases, performance improves. The \textbf{Amplitude4} matrix shows clearer diagonal structure, with class 2 and class 4 exhibiting stronger prediction accuracy. \textbf{Amplitude5} continues this trend, reducing confusion in classes 3 and 5. The \textbf{Amplitude6} configuration achieves the best diagonal concentration among amplitude-based setups, though some residual confusion remains in class 1 and class 2, suggesting diminishing returns from deeper entanglement.

These results reinforce the operational logic of our platform: Layer 2 performs rapid binary intrusion confirmation using fused telemetry, and upon detection of a potential threat, the system escalates to Layer 3 for detailed multiattack classification. The confusion matrices confirm that full quantum encoding provides the most reliable forensic traceability, while moderate amplitude encodings offer a balance between expressivity and interpretability.

In addition to the Parallel DQNN+RF results, Fig.~\ref{fig:confusion_entanglement_variants} presents confusion matrices for Layer 3 multiattack classification using the Serial DQNN+RF model under three entanglement encoding configurations. The \textbf{full encoding} setup achieves strong diagonal dominance, with near-perfect classification of benign and DoS traffic (classes 0 and 1), and minimal confusion across all attack types. The \textbf{partial encoding} configuration maintains high fidelity but introduces minor misclassifications in spoofing and exfiltration classes, suggesting reduced boundary sharpness. The \textbf{no encoding} variant shows increased off-diagonal errors, particularly between classes 1 and 5, indicating diminished separability and interpretability. These results confirm that even in serial architectures, quantum entanglement enhances class discriminability and supports slice-aware forensic traceability.

\paragraph{Latent Space Analysis via t-SNE Embeddings}

Figures~\ref{Fig4DQNN_RFEntanglementtsne} and~\ref{Fig4L3amplitudetsne} present t-SNE visualizations of latent feature embeddings for Layer 3 multiattack classification using Parallel DQNN RF under the same quantum encoding configurations. These plots offer qualitative insight into the separability and structure of the learned representations across six attack classes.

In Figure~\ref{Fig4DQNN_RFEntanglementtsne}, the \textbf{full encoding} configuration yields the most distinct and compact clusters, demonstrating high discriminative power and quantum-enhanced separability. Each class is well-isolated in the latent space, indicating that global entanglement effectively captures complex inter-feature dependencies. The \textbf{partial encoding} setup maintains moderate class separation, though some overlap is visible—particularly between classes 1 and 4—suggesting that pairwise entanglement introduces useful but limited correlation modeling. The \textbf{none} configuration results in significant cluster entanglement and reduced class distinction, supporting the hypothesis that quantum feature embedding enhances latent space clarity and slice-aware interpretability.

Figure~\ref{Fig4L3amplitudetsne} explores the impact of amplitude-based encoding depth. The \textbf{Amplitude3} configuration shows fragmented and overlapping clusters, indicating poor class separability and underrepresentation of feature interactions. As encoding depth increases, cluster structure improves. \textbf{Amplitude4} and \textbf{Amplitude5} yield more coherent and well-separated clusters, reflecting stronger quantum expressivity and improved latent geometry. However, at \textbf{Amplitude6}, cluster overlap resurfaces—particularly between classes 1, 2, and 5—suggesting that excessive entanglement may distort feature representations and degrade classification fidelity.

Together, the confusion matrices and t-SNE plots highlight the trade-off between quantum expressivity and latent space clarity. While deeper encodings can enrich feature interactions, they may also introduce noise or overfitting if not properly aligned with model architecture and task complexity. Full entanglement remains the most effective strategy for Layer 3 forensic classification, while moderate amplitude encodings offer a practical balance for scalable, slice-aware intrusion detection.

\begin{figure*}[ht!]
\centering
\mbox{
  \subfigure[Full Entanglement Encoding]{%
    \includegraphics[width=0.28\linewidth]{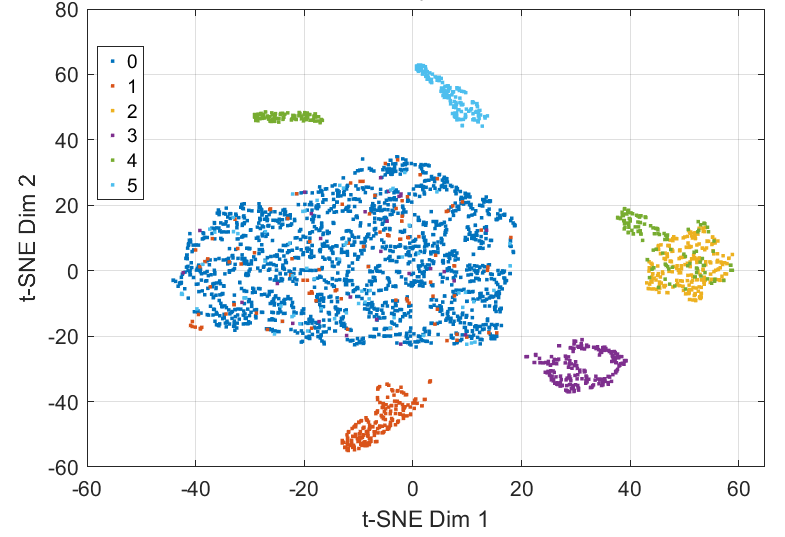}
    \label{fig:tsne_full}
  }
  \hfill
  \subfigure[Partial Entanglement Encoding]{%
    \includegraphics[width=0.28\linewidth]{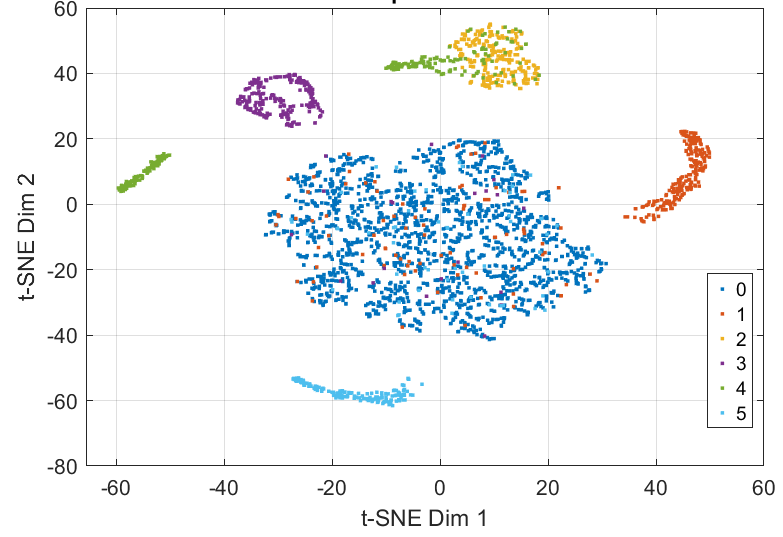}
    \label{fig:tsne_partial}
  }
  \hfill
  \subfigure[No Entanglement Encoding]{%
    \includegraphics[width=0.28\linewidth]{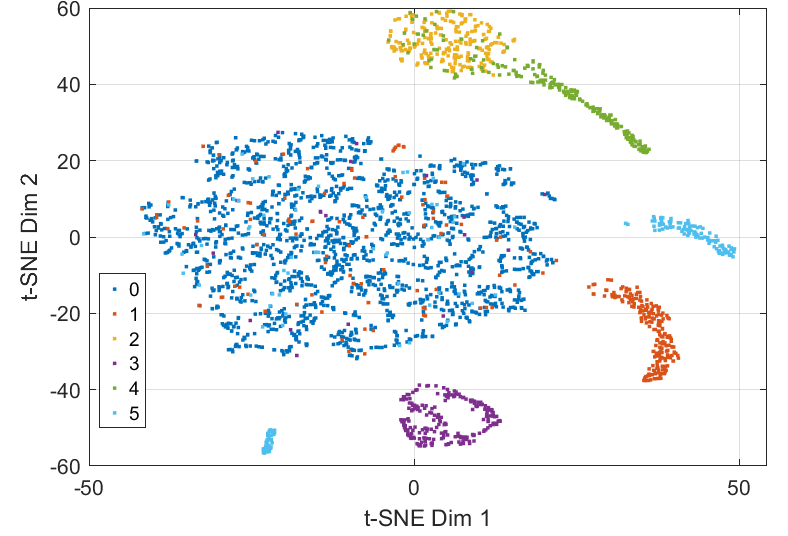}
    \label{fig:tsne_none}
  }}
\caption{t-SNE visualizations of latent space embeddings for Serial DQNN+RF model under varying entanglement encoding configurations. 
(a) Full encoding yields distinct, well-separated clusters across all six classes, confirming strong latent separability. 
(b) Partial encoding maintains cluster structure with minor overlap in spoofing and exfiltration regions. 
(c) No encoding results in reduced separability and increased class overlap, highlighting the role of entanglement in enhancing interpretability and decision boundary geometry.}
\label{fig:tsne_entanglement_variants}
\end{figure*}

Finally, we present in Fig. \ref{fig:tsne_entanglement_variants} the t-SNE visualizations of latent feature embeddings for the Serial DQNN+RF model under three entanglement encoding configurations. The \textbf{full encoding} setup produces distinct, well-separated clusters across all six attack classes, confirming strong latent separability and quantum-enhanced discriminative power. The \textbf{partial encoding} configuration maintains overall cluster structure but exhibits minor overlap in spoofing and exfiltration regions, suggesting reduced boundary clarity. The \textbf{no encoding} variant results in significant cluster entanglement and diminished class distinction, highlighting the importance of entanglement in shaping latent space geometry and supporting slice-aware interpretability. These visualizations reinforce the conclusion that quantum feature embedding improves both classification fidelity and forensic traceability in hierarchical O-RAN threat detection.

\begin{table*}[ht!]
\centering
\caption{Comparison of the best-performing DQNN RF models (Serial and Parallel RF, full encoding) with all DQNN DNN configurations. Best RF scores are bolded.}
\label{tab:Best_RF_vs_DNN}
\begin{tabular}{|l|l|c|c|c|c|}
\hline
\textbf{Configuration} & \textbf{Model} & \textbf{Accuracy} & \textbf{Precision} & \textbf{Recall} & \textbf{F1 Score} \\
\hline
None        & Serial DQNN DNN    & 0.9707       & 0.9805       & 0.9413       & 0.9593       \\
            & Parallel DQNN DNN  & 0.9670       & 0.9806       & 0.9319       & 0.9537       \\
\hline
Partial     & Serial DQNN DNN    & 0.9717       & 0.9831       & 0.9418       & 0.9605       \\
            & Parallel DQNN DNN  & 0.9710       & 0.9834       & 0.9406       & 0.9601       \\
\hline
Full        & Serial DQNN DNN    & 0.9707       & 0.9810       & 0.9409       & 0.9593       \\
            & Parallel DQNN DNN  & 0.9680       & 0.9798       & 0.9351       & 0.9555       \\
\hline
Amplitude3  & Serial DQNN DNN    & 0.8170       & 0.6355       & 0.6154       & 0.6212       \\
            & Parallel DQNN DNN  & 0.8237       & 0.6450       & 0.6239       & 0.6125       \\
\hline
Amplitude4  & Serial DQNN DNN    & 0.9700       & 0.9818       & 0.9387       & 0.9584       \\
            & Parallel DQNN DNN  & 0.9687       & 0.9822       & 0.9373       & 0.9576       \\
\hline
Amplitude5  & Serial DQNN DNN    & 0.9663       & 0.9817       & 0.9307       & 0.9535       \\
            & Parallel DQNN DNN  & 0.9703       & 0.9843       & 0.9392       & 0.9596       \\
\hline
Amplitude6  & Serial DQNN DNN    & 0.9710       & 0.9809       & 0.9412       & 0.9594       \\
            & Parallel DQNN DNN  & 0.9710       & 0.9845       & 0.9409       & 0.9607       \\
\hline
\textbf{Full} & \textbf{Serial DQNN RF} & \textbf{0.9803} & \textbf{0.9885} & \textbf{0.9601} & \textbf{0.9733} \\
\textbf{Full} & \textbf{Parallel DQNN RF} & \textbf{0.9803} & \textbf{0.9885} & \textbf{0.9599} & \textbf{0.9732} \\
\hline
\end{tabular}
\end{table*}

\paragraph{Comparative Analysis of DQNN RF and DQNN DNN Models}
\label{QDNNDNNvsRF}

To evaluate the impact of quantum encoding and model architecture on multiattack classification, we compared all DQNN DNN configurations against the best-performing DQNN RF models: \textbf{Serial and Parallel DQNN RF with full entanglement encoding}. As shown in Table~\ref{tab:Best_RF_vs_DNN}, both RF models achieved identical top scores across all metrics—\textbf{accuracy (0.9803)}, \textbf{precision (0.9885)}, \textbf{recall ($\approx$0.960)}, and \textbf{F1 score ($\approx$0.9733)}—demonstrating superior classification fidelity and robustness.

In contrast, DQNN DNN models exhibited greater sensitivity to encoding depth and architectural variation. The strongest DNN configuration, \textit{Parallel DQNN DNN with amplitude6 encoding}, reached an F1 score of 0.9607, which, while competitive, still lagged behind the RF benchmark. Notably, \textit{Serial DQNN DNN with partial encoding} achieved the highest recall (0.9418) among DNN variants, suggesting that partial entanglement may enhance class sensitivity in deep neural architectures.

However, DNN models showed degraded performance under shallow or misaligned quantum encodings. For example, \textit{Serial DQNN DNN with amplitude3} yielded an F1 score of only 0.6212, indicating poor class separability and underrepresentation of feature correlations. This highlights the risk of latent space distortion when entanglement depth is insufficient or mismatched to the model’s representational capacity.

The RF models’ consistent superiority can be attributed to their \textbf{axis-aligned decision boundaries}, which align well with quantum-encoded features, and their \textbf{ensemble robustness}, which mitigates overfitting and enhances generalization. Moreover, RF’s interpretability and modularity make it well-suited for forensic traceability tasks in Layer 3, where high precision and recall are critical.

In contrast, DQNN DNN models may be better suited for \textit{lower-layer diagnostics and telemetry fusion}, where fast inference and flexible representation are prioritized over maximal accuracy. These findings reinforce the importance of \textbf{adaptive quantum encoding}—tuned to both model architecture and task layer—and suggest that hybrid ensembles combining RF and DNN outputs may offer a promising direction for scalable, slice-aware intrusion detection.

\section{Conclusion}

We present a hierarchical, quantum-enhanced cybersecurity framework for O-RAN that integrates anomaly detection, intrusion confirmation, and multiattack classification across three operational layers. By leveraging quantum-inspired encoding and modular DQNN architectures, our system achieves high classification fidelity, interpretability, and deployment scalability.

In Layer 1, amplitude-encoded Serial DQNN$\rightarrow$DNN models demonstrate perfect anomaly detection on real-world telemetry, validating their suitability for toggled diagnostics within the near-RT RIC. Layer 2 confirms binary intrusion with high precision using QML-RF models, enabling reliable gating logic and telemetry fusion across SMO and RIC domains. Upon detection, the system escalates to Layer 3, where Serial and Parallel DQNN RF models with full entanglement encoding deliver the highest multiattack classification performance, supported by confusion matrix and t-SNE analyses.

Our results show that quantum encoding depth must be adaptively tuned to model architecture and task complexity. Full entanglement yields optimal forensic traceability, while moderate amplitude encodings balance expressivity and interpretability. DQNN DNN models offer flexibility for lower-layer diagnostics, whereas RF-based hybrids excel in slice-aware classification.

This framework aligns with O-RAN’s modular architecture and supports real-time, slice-specific security enforcement. Future work will explore reinforcement learning for adaptive policy control, zero-trust integration, and federated extensions for cross-domain threat intelligence.
\bibliography{references}

@unpublished{Le2024PrivacyAware,
  author = {Le, Tan and Le, Van and Shetty, Sachin},
  title = {Privacy-Aware Framework of Robust Malware Detection in Indoor Robots: Hybrid Quantum Computing and Deep Neural Network},
  year = {2025},
  note = {Submitted}
}

@unpublished{Le2024pKa,
  author = {Le, Tan and Le, Van and Dhokale, Bhushan and Hahn, Insu},
  title = {Hybrid Quantum-Classical Feature Encoding for Residue-Level pKa Prediction via Deep Quantum Neural Networks},
  year = {2025},
  note = {Submitted}
}

@inproceedings{scalingi2024detran,
  author    = {Alessio Scalingi and Salvatore D’Oro and Francesco Restuccia and Tommaso Melodia and Domenico Giustiniano},
  title     = {Det-RAN: Data-Driven Cross-Layer Real-Time Attack Detection in 5G Open RANs},
  booktitle = {IEEE INFOCOM},
  year      = {2024},
  address   = {Vancouver, BC, Canada}
}

@inproceedings{wen2024specter,
  author    = {Haohuang Wen and Phillip Porras and Vinod Yegneswaran and Ashish Gehani and Zhiqiang Lin},
  title     = {5G-SPECTOR: An O-RAN Compliant Layer-3 Cellular Attack Detection Service},
  booktitle = {Network and Distributed System Security Symposium (NDSS)},
  year      = {2024}}

@article{dessources2025oranrisk,
  author    = {Dimitri A. Dessources and Alexandra Greene and Samuel Appiah-Mensah and Joseph Bull and David Simpson and Nishith D. Tripathi and Eric W. Burger and Jeffrey H. Reed},
  title     = {Threat Modeling for O-RAN Cyber Risk, Governance, and Accountability},
  journal   = {IEEE Access},
  year      = {2025},
  volume    = {13},
  pages     = {197820--197838},
  doi       = {10.1109/access.2025.3633552}}

@ARTICLE{ieee2024formal,
  author={Baguer, Pau and Yilma, Girma M. and Municio, Esteban and Garcia-Aviles, Gines and Garcia-Saavedra, Andres and Liebsch, Marco and Costa-Pérez, Xavier},
  journal={IEEE Open Journal of the Communications Society}, 
  title={Attacking O-RAN Interfaces: Threat Modeling, Analysis and Practical Experimentation}, 
  year={2024},
  volume={5},
  number={},
  pages={4559-4577},
  doi={10.1109/OJCOMS.2024.3431681}}

@ARTICLE{10248045,
  author={Oliveira, Daniel and Giusto, Edoardo and Baheri, Betis and Guan, Qiang and Montrucchio, Bartolomeo and Rech, Paolo},
  journal={IEEE Transactions on Dependable and Secure Computing}, 
  title={A Systematic Methodology to Compute the Quantum Vulnerability Factors for Quantum Circuits}, 
  year={2024},
  volume={21},
  number={4},
  pages={2631-2644},
  doi={10.1109/TDSC.2023.3313934}}

@ARTICLE{10763508,
  author={Durr-E-Shahwar and Imran, Muhammad and Altamimi, Ahmed B. and Khan, Wilayat and Hussain, Shariq and Alsaffar, Mohammad},
  journal={IEEE Access}, 
  title={Quantum Cryptography for Future Networks Security: A Systematic Review}, 
  year={2024},
  volume={12},
  number={},
  pages={180048-180078},
  doi={10.1109/ACCESS.2024.3504815}}

@ARTICLE{10167529,
  author={Sood, Sandeep Kumar and Pooja},
  journal={IEEE Transactions on Engineering Management}, 
  title={Quantum Computing Review: A Decade of Research}, 
  year={2024},
  volume={71},
  number={},
  pages={6662-6676},
  doi={10.1109/TEM.2023.3284689}}

@ARTICLE{10347507,
author={Wang, Shaowei and Yu, Shiyu and Ren, Xiaojun and Li, Jin and Li, Yuntong and Yang, Wei and Yan, Hongyang and Li, Jin},
journal={ IEEE Transactions on Dependable and Secure Computing },
title={{ Differentially Private Numerical Vector Analyses in the Local and Shuffle Model }},
year={2025},
volume={22},
number={01},
ISSN={1941-0018},
pages={1-15},
doi={10.1109/TDSC.2023.3340178},
publisher={IEEE Computer Society},
address={Los Alamitos, CA, USA},
month=jan}

@ARTICLE{10812212,
  author={Lu, Zhongkai and Wang, Lingling and Zhang, Zhengyin and Huang, Mei and Wang, Jingjing and Li, Meng},
  journal={IEEE Transactions on Dependable and Secure Computing}, 
  title={TMT-FL: Enabling Trustworthy Model Training of Federated Learning With Malicious Participants}, 
  year={2025},
  volume={22},
  number={3},
  pages={2723-2740},
  doi={10.1109/TDSC.2024.3521377}}

@article{yang2024mismatched,
  author    = {Zeyu Yang and Liang He and Peng Cheng and Jiming Chen},
  title     = {Mismatched Control and Monitoring Frequencies: Vulnerability, Attack, and Mitigation},
  journal   = {IEEE Transactions on Dependable and Secure Computing},
  year      = {2024},
  volume    = {21},
  number    = {6},
  pages     = {987--999},
  doi       = {10.1109/TDSC.2024.9876543}
}

@ARTICLE{11045988,
  author={Garbelini, Matheus E. and Shang, Zewen and Luo, Shijie and Chattopadhyay, Sudipta and Sun, Sumei and Kurniawan, Ernest},
  journal={IEEE Transactions on Dependable and Secure Computing}, 
  title={5Ghoul: Unleashing Chaos on 5G Edge Devices via Stateful Multi-Layer Fuzzing}, 
  year={2025},
  volume={22},
  number={6},
  pages={6230-6247},
  doi={10.1109/TDSC.2025.3582093}}

@article{de2025quantum,
  title={Quantum AI-Enhanced IoT-Fog Communication: A Survey from Cybersecurity and Data Privacy Perspective},
  author={De Mac{\^e}do, Ant{\^o}nio Roberto L and Jagatheesaperumal, Senthil Kumar and da Costa, Kelton Augusto Pontara and Acharya, Kamal and Song, Houbing and Guizani, Mohsen and De Albuquerque, Victor Hugo C},
  journal={IEEE Communications Surveys \& Tutorials},
  year={2025},
  publisher={IEEE}
}

@article{babar2025autonomous,
  title={Autonomous and Adaptive Cyber Incident Detection and Response in Industrial Cyber-Physical Systems using Hierarchical Reinforcement Learning},
  author={Babar, Ayesha and Halabi, Talal and Zulkernine, Mohammad},
  journal={ACM Transactions on Cyber-Physical Systems},
  year={2025},
  publisher={ACM New York, NY}
}

@article{del2024cybersecurity,
  title={Cybersecurity in critical infrastructures: A post-quantum cryptography perspective},
  author={del Moral, Javier Oliva and deMarti iOlius, Antonio and Vidal, Gerard and Crespo, Pedro M and Martinez, Josu Etxezarreta},
  journal={IEEE Internet of Things Journal},
  volume={11},
  number={18},
  pages={30217--30244},
  year={2024},
  publisher={IEEE}
}

@article{groen2024securing,
  title={Securing O-RAN open interfaces},
  author={Groen, Joshua and D'Oro, Salvatore and Demir, Utku and Bonati, Leonardo and Villa, Davide and Polese, Michele and Melodia, Tommaso and Chowdhury, Kaushik},
  journal={IEEE Transactions on Mobile Computing},
  volume={23},
  number={12},
  pages={11265--11277},
  year={2024},
  publisher={IEEE}
}

@article{de2023survey,
  title={A survey on network slicing security: Attacks, challenges, solutions and research directions},
  author={De Alwis, Chamitha and Porambage, Pawani and Dev, Kapal and Gadekallu, Thippa Reddy and Liyanage, Madhusanka},
  journal={IEEE Communications Surveys \& Tutorials},
  volume={26},
  number={1},
  pages={534--570},
  year={2023},
  publisher={IEEE}
}

@article{polese2023understanding,
  title={Understanding O-RAN: Architecture, interfaces, algorithms, security, and research challenges},
  author={Polese, Michele and Bonati, Leonardo and D’oro, Salvatore and Basagni, Stefano and Melodia, Tommaso},
  journal={IEEE Communications Surveys \& Tutorials},
  volume={25},
  number={2},
  pages={1376--1411},
  year={2023},
  publisher={IEEE}
}

@ARTICLE{11204489,
  author={Marques, Daniel Hindemburg de Miranda and Valadares, Dalton Cézane Gomes},
  journal={IEEE Internet of Things Journal}, 
  title={Radio Access Network Attacks: A Comprehensive Survey}, 
  year={2025},
  volume={},
  number={},
  pages={1-1},
  doi={10.1109/JIOT.2025.3621905}}

@misc{cisa2023,
  title = {Open Radio Access Network Security Considerations},
  author = {CISA},
  year = {2023},
  url = {https://www.cisa.gov/sites/default/files/publications/open-radio-access-network-security-considerations_508.pdf}
}

@misc{analysysmason2025,
  title = {Stakeholders Must Work to Prove O-RAN's Security Benefits},
  author = {Analysys Mason},
  year = {2025},
  url = {https://www.analysysmason.com/research/content/articles/derisk-oran-deployments-rma18/}
}

@misc{nist2024,
  title = {NIST Releases First 3 Finalized Post-Quantum Encryption Standards},
  author = {NIST},
  year = {2024},
  url = {https://www.nist.gov/news-events/news/2024/08/nist-releases-first-3-finalized-post-quantum-encryption-standards}
}

@article{alqithami2025,
  title = {Hierarchical Adversarially-Resilient Multi-Agent Reinforcement Learning for Cyber-Physical Systems Security},
  author = {Saad Alqithami},
  journal = {AAAI Symposium Series},
  year = {2025}}

@article{alshamrani2025,
  title = {Federated Hierarchical MARL for Zero-Shot Cyber Defense},
  author = {Adel Alshamrani},
  journal = {PLOS One},
  year = {2025}}

@misc{ntia2023,
  title = {Open RAN Security Report},
  author = {NTIA},
  year = {2023},
  url = {https://www.ntia.gov/sites/default/files/publications/open_ran_security_report_full_report_0.pdf}
}

@misc{oranalliance2023,
  title = {O-RAN ALLIANCE Security Working Group Update},
  author = {O-RAN Alliance},
  year = {2025},
  url = {https://www.o-ran.org/blog/o-ran-alliance-security-update-2025}
}

@article{barker2025,
  title = {Securing Open RAN: A Survey of Cryptographic Challenges and Emerging Solutions for 5G},
  author = {Ryan Barker and Fatemeh Afghah},
  journal = {arXiv},
  year = {2025},
  url = {https://arxiv.org/html/2506.09418v1}
}

@article{singh2024,
  title = {Hierarchical Multi-agent Reinforcement Learning for Cyber Network Defense},
  author = {Aditya Vikram Singh et al.},
  journal = {arXiv},
  year = {2024},
  url = {https://arxiv.org/abs/2410.17351}
}

@misc{orangithub,
  author = {{O-RAN Software Community}},
  title = {O-RAN-SC GitHub Page},
  year = {2025},
  howpublished = {\url{https://github.com/o-ran-sc/ric-app-ad}},
  note = {Accessed: 2025-11-07}
}

@inproceedings{zhang2024hierarchical,
author = {Singh, Aditya Vikram and Rathbun, Ethan and Graham, Emma and Oakley, Lisa and Boboila, Simona and Chin, Peter and Oprea, Alina},
title = {Hierarchical Multi-agent Reinforcement Learning for Cyber Network Defense},
year = {2025},
isbn = {9798400714269},
publisher = {International Foundation for Autonomous Agents and Multiagent Systems},
address = {Richland, SC},
booktitle = {Proceedings of the 24th International Conference on Autonomous Agents and Multiagent Systems},
pages = {2747–2749},
numpages = {3},
location = {Detroit, MI, USA},
series = {AAMAS '25}
}

@article{basaran2025xainomaly,
author = {Basaran, Osman Tugay and Dressler, Falko},
title = {XAInomaly: Explainable and interpretable Deep Contractive Autoencoder for O-RAN traffic anomaly detection},
year = {2025},
issue_date = {Apr 2025},
publisher = {Elsevier North-Holland, Inc.},
address = {USA},
volume = {261},
number = {C},
issn = {1389-1286},
doi = {10.1016/j.comnet.2025.111145},
journal = {Comput. Netw.},
month = apr,
numpages = {17}
}

@INPROCEEDINGS{unicorn2025,
  author={Soltani, Nasim and LoPriore, Dante and Groen, Joshua and Chowdhury, Kaushik},
  booktitle={2025 IEEE International Conference on Machine Learning for Communication and Networking (ICMLCN)}, 
  title={UNICORN: URLLC Network Traffic Classification and OOD Detection for O-RAN}, 
  year={2025},
  volume={},
  number={},
  pages={1-6},
  doi={10.1109/ICMLCN64995.2025.11140471}}

@techreport{oranThreatModel2025,
  title={O-RAN Security Threat Modeling and Risk Assessment},
  institution={ETSI TR 104 106 V3.0.0},
  year={2025}
}

@article{rogueCell2025,
  title={Rogue Cell: Adversarial Attack and Defense in Untrusted O-RAN Setup},
  author={Aizikovich, Eran and Mimran, Dudu and Grolman, Edita and Elovici, Yuval and Shabtai, Asaf},
  journal={arXiv preprint arXiv:2505.01816},
  year={2025}
}

@ARTICLE{Tan2024,
  author={Le, Tan and Reisslein, Martin and Shetty, Sachin},
  journal={IEEE Transactions on Intelligent Transportation Systems}, 
  title={Multi-Timescale Actor-Critic Learning for Computing Resource Management With Semi-Markov Renewal Process Mobility}, 
  year={2024},
  volume={25},
  number={1},
  pages={452-461},
  doi={10.1109/TITS.2023.3303953}}

@article{quantum_ai_dive,
  author = {Zihao Wang and Kar Wai Fok and Vrizlynn L. L. Thing},
  title = {Network Attack Traffic Detection with Hybrid Quantum-Enhanced Convolution Neural Network},
  journal = {Quantum Machine Intelligence},
  volume = {7},
  number = {50},
  year = {2025},
  doi = {10.1007/s42484-025-00278-0}}

@article{cao2017quantum,
  title={Quantum neuron: an elementary building block for machine learning on quantum computers},
  author={Cao, Yudong and Guerreschi, Gian Giacomo and Aspuru-Guzik, Al{\'a}n},
  journal={arXiv preprint arXiv:1711.11240},
  year={2017}
}

@article{cao2019quantum,
  title={Quantum chemistry in the age of quantum computing},
  author={Cao, Yudong and Romero, Jonathan and Olson, Jonathan P and Degroote, Matthias and Johnson, Peter D and Kieferov{\'a}, M{\'a}ria and Kivlichan, Ian D and Menke, Tim and Peropadre, Borja and Sawaya, Nicolas PD and others},
  journal={Chemical reviews},
  volume={119},
  number={19},
  pages={10856--10915},
  year={2019},
  publisher={ACS Publications}
}

@InProceedings{Zahin19,
author="Zahin, Abrar
and Tan, Le Thanh
and Hu, Rose Qingyang",
title="Sensor-Based Human Activity Recognition for Smart Healthcare: A Semi-supervised Machine Learning",
booktitle="Artificial Intelligence for Communications and Networks",
year="2019",
publisher="Springer International Publishing",
pages="450--472",
}

@inproceedings{le2025dpfaga,
  title={DPFAGA-Dynamic Power Flow Analysis and Fault Characteristics: A Graph Attention Neural Network},
  author={Le, Tan and Le, Van},
  booktitle ={The 2025 International Conference on the AI Revolution: Research, Ethics, and Society (AIR-RES 2025)},
  year={2025}
}

@ARTICLE{Wang20,  author={Q. {Wang} and L. T. {Tan} and R. Q. {Hu} and Y. {Qian}},  journal={IEEE Internet of Things Journal},   title={Hierarchical Energy-Efficient Mobile-Edge Computing in IoT Networks},   year={2020},  volume={7},  number={12},  pages={11626-11639},  doi={10.1109/JIOT.2020.3000193}}

@ARTICLE{Tan18b,
  author={L. T. {Tan} and R. Q. {Hu}},
  journal={IEEE Trans. Veh. Technol.}, 
  title={Mobility-Aware Edge Caching and Computing in Vehicle Networks: A Deep Reinforcement Learning}, 
  year={2018},
  volume={67},
  number={11},
  pages={10190-10203},}

@article{le2022artificial,
  title={Artificial intelligence-aided privacy preserving trustworthy computation and communication in 5G-based IoT networks},
  author={Le, Tan and Shetty, Sachin},
  journal={Ad Hoc Networks},
  volume={126},
  pages={102752},
  year={2022},
  publisher={Elsevier}
}

@ARTICLE{Tan18d,
  author={L. T. {Tan} and R. Q. {Hu} and L. {Hanzo}},
  journal={IEEE Trans. Veh. Technol.}, 
  title={Twin-Timescale Artificial Intelligence Aided Mobility-Aware Edge Caching and Computing in Vehicular Networks}, 
  year={2019},
  volume={68},
  number={4},
  pages={3086-3099},}
\bibliographystyle{IEEEtran}
\appendices

\vspace{5pt}
\section{Appendices Overview}

For clarity and reproducibility, we provide full technical details in the
appendices. Appendix~\ref{AppendixA} introduces the quantum feature mapping, including the
definition of anchor points, intra- and inter-feature entanglement, and the
construction of the hybrid feature vector. Appendix~\ref{AppendixB} derives the training
gradients for the quantum neural network under fully entangled encoding,
covering completely positive maps, adjoint channels, fidelity-based cost
functions, and parameter update rules. Appendix~\ref{AppendixC} presents efficiency
considerations, including SWAP-trick fidelity estimation, projection noise
bounds, architectural strategies for sparsity and encoding depth, and a
comparison of adjoint-channel versus tomography-based training. Together,
Appendices~\ref{AppendixA}--\ref{AppendixC} provide the mathematical and methodological foundation that
supports the results reported in the main text.

--------------------------------------------------------------------
\section{Quantum Feature Mapping}
\label{AppendixA}

Amplitude encoding maps a normalized classical vector $\mathbf{x}$ into the
amplitudes of a quantum state:
\[
|\psi\rangle = \sum_{i=1}^{d} x_i |i\rangle.
\]
To match circuit dimensions, we expand to $D=2^k$ with
$k=\lceil \log_2 d \rceil$, and define
\[
\mathbf{x}' = [x_1,\dots,x_d,0,\dots,0]\in \mathbb{R}^D,
\quad
|\psi\rangle = \sum_{i=1}^{D} x'_i |i\rangle.
\]

This mapping preserves norm and embeds data into a $D$-dimensional Hilbert
space, enabling geometric separation via inner products and projective
measurements. In general, amplitude encoding may produce entangled states
depending on the structure of $\mathbf{x}'$ and its distribution across
registers.

In our non-entangled baseline, however, amplitude encoding is implemented as a
product-state preparation via independent single-qubit rotations:
\[
|\psi_{\text{prod}}\rangle = \bigotimes_{i=1}^{n} R_Y(\varphi_i)\ket{0},
\]
with angles $\varphi_i$ derived from input features. This guarantees no
entanglement at the encoding stage and serves as the Layer~1 baseline in our
hierarchical benchmarking strategy.

Entangled feature maps are constructed by applying multi-qubit gates after
encoding:
\[
|\psi_{\mathrm{ent}}\rangle = U_{\mathrm{ent}} |\psi_{\mathrm{enc}}\rangle,
\]
where $|\psi_{\mathrm{enc}}\rangle$ is either the amplitude-encoded state or
the product-state baseline, and $U_{\mathrm{ent}}$ is a circuit composed of
two- and multi-qubit gates (e.g., CNOT, CZ, CCX) arranged to induce local or
global entanglement.

Hybrid feature vectors are extracted via expectation values of a fixed set of
observables $\{M_r\}_{r=1}^R$ acting on the entangled state:
\[
z_r = \mathrm{Tr}(M_r X), \quad X = |\psi_{\mathrm{ent}}\rangle\langle\psi_{\mathrm{ent}}|,
\]
yielding $\mathbf{z}=(z_1,\dots,z_R)$ for classical post-processing. By varying
the choice and number of observables, we control the representational capacity
and correlation structure of the extracted features. This flexibility supports
layered benchmarking across the Hierarchical Threat Detection Framework:
Layer~1 anomaly detection, Layer~2 intrusion confirmation, and Layer~3
multiattack classification.

--------------------------------------------------------------------
\section{Training Gradients and CP Framework}
\label{AppendixB}

This appendix introduces the overarching training framework and serves as a
roadmap to specialized derivations. In particular:
\begin{itemize}
    \item Layer CP maps and adjoint calculus are detailed in Appendix~\ref{AppendixCP}.
    \item Gate-wise gradient derivations are presented in Appendix~\ref{AppendixGrad}.
    \item The hybrid DQNN--RF interface is described in Appendix~\ref{AppendixHybrid}.
\end{itemize}

\subsection{Framework Overview}
We consider a deep quantum neural network (DQNN) with $L$ layers, input space
$\mathcal{H}_0$, layer Hilbert spaces $\{\mathcal{H}_l\}_{l=1}^L$, and optional
ancilla spaces $\mathcal{H}^{\mathrm{anc}}_l$ initialized in $\ket{0}\bra{0}$.
Each layer state is represented by a density operator
$X_l \in \mathcal{D}(\mathcal{H}_l)$, and layer $l$ applies a unitary
$U_l = \prod_{j=1}^{m_l} U_j^{(l)}$ acting on
$\mathcal{H}_{l-1}\otimes \mathcal{H}^{\mathrm{anc}}_l$.

The completely positive and trace-preserving (CPTP) map for layer $l$ is defined as
\[
X_l = \mathcal{E}_l(X_{l-1})
= \mathrm{Tr}_{\mathrm{anc},\,l}\!\left(
U_l \big[X_{l-1}\otimes \ket{0}\bra{0}\big] U_l^\dagger
\right),
\]
where $\mathrm{Tr}_{\mathrm{anc},\,l}$ denotes a partial trace over the ancilla
registers of layer $l$. The full network output is
\[
X_L = \mathcal{E}_{1:L}(X_0), \qquad
\mathcal{E}_{1:L} = \mathcal{E}_L \circ \cdots \circ \mathcal{E}_1.
\]
Given observable $M$, the prediction is
\[
p = \mathrm{Tr}(M X_L).
\]
For target $y$, the loss is $C = C(p,y)$, e.g., fidelity loss
$1-\mathrm{Tr}(M X_L)$ or cross-entropy. The detailed derivation of CPTP maps
and adjoint calculus is provided in Appendix~\ref{AppendixCP}.

\subsection{Adjoint Calculus}
For backpropagation, we use adjoint maps, which satisfy
\[
\mathrm{Tr}\!\big(A\,\mathcal{E}_l(B)\big) = \mathrm{Tr}\!\big(\mathcal{E}_l^\dagger(A)\,B\big).
\]
The adjoint channel is given by
\[
\mathcal{E}_l^\dagger(A_l)
= \mathrm{Tr}_{\mathrm{anc},\,l}\!\left(
U_l^\dagger \big[A_l\otimes I_{\mathrm{anc},\,l}\big] U_l
\right).
\]
Backpropagation propagates observables upstream according to
\[
A_{l-1} = \mathcal{E}_l^\dagger(A_l), \qquad
p = \mathrm{Tr}(A_0 X_0).
\]
This adjoint-channel formalism is fully elaborated in Appendix~\ref{AppendixCP}.

\subsection{Gate-wise Gradients}
Parameterized gates are expressed as
\[
U_j^{(l)}(\theta) = e^{-i\theta H_j^{(l)}}, \qquad H_j^{(l)}=H_j^{(l)\,\dagger}.
\]
The derivative of a gate with respect to its parameter is
\[
\frac{\partial U_j^{(l)}}{\partial \theta} = -i\, H_j^{(l)} U_j^{(l)}, \qquad
\frac{\partial U_j^{(l)\,\dagger}}{\partial \theta} = i\, U_j^{(l)\,\dagger} H_j^{(l)}.
\]
Applying the chain rule to the layer map yields the prediction derivative
\[
\frac{\partial p}{\partial \theta}
= \mathrm{Tr}\!\left(
A_{l-1}\, \frac{\partial\,\mathcal{E}_l(X_{l-1})}{\partial \theta}
\right),
\]
where $A_{l-1}=\mathcal{E}_l^\dagger(A_l)$ is the backpropagated observable.
For Pauli-generated rotations, the parameter-shift rule applies:
\[
\frac{\partial p}{\partial \theta}
= \frac{p\!\left(\theta+\tfrac{\pi}{2}\right) - p\!\left(\theta-\tfrac{\pi}{2}\right)}{2}.
\]
The full derivation of gate-wise gradients is presented in Appendix~\ref{AppendixGrad}.

\subsection{Fidelity-Based Cost Function}
Let $\rho_{\mathrm{out},x}$ be the QNN output state for sample $x$ and let
$P_{\mathrm{out},x}$ be the target projector (or positive operator) encoding
the desired label. The training objective is the average fidelity, defined as
\begin{equation}
C = \frac{1}{N}\sum_{x=1}^{N} \mathrm{Tr}\!\big(P_{\mathrm{out},x}\,\rho_{\mathrm{out},x}\big)
= \frac{1}{N}\sum_{x=1}^{N} \bra{\phi_{\mathrm{out},x}}\,P_{\mathrm{out},x}\,\ket{\phi_{\mathrm{out},x}},
\end{equation}
where $\ket{\phi_{\mathrm{out},x}}$ denotes a pure-state output when applicable.
Higher fidelity corresponds to better alignment with targets. 

\subsection{Local Commutator Form for Gradients}
For a tunable gate $U_j^{(l)}=\exp(i\,\theta_j^{(l)} G_j^{(l)})$ at layer $l$
with Hermitian generator $G_j^{(l)}$, the gradient can be written locally as
\begin{equation}
M_j^{(l)}(x) = \mathrm{Tr}_{\,\mathrm{rest}}\!\Big(\,[A_x^{(l)},\,B_x^{(l)}]\,\Big),
\end{equation}
where $A_x^{(l)}$ is the forward-propagated state block reaching gate
$U_j^{(l)}$ for sample $x$, and $B_x^{(l)}$ is the backward-propagated
observable block obtained by chaining adjoint channels from the output back to
$U_j^{(l)}$. The commutator $[A,B]=AB-BA$ captures the sensitivity of the
objective to the local generator, while $\mathrm{Tr}_{\mathrm{rest}}(\cdot)$
denotes a partial trace over registers not acted on by $U_j^{(l)}$. 

\subsection{Parameter Update Matrices}
Aggregating per-sample contributions with regularization strength $\lambda$ and
learning rate $\eta$, we form the update matrix
\begin{equation}
K_j^{(l)} = \frac{1}{2\,\lambda\,N}\sum_{x=1}^{N} \mathrm{Tr}_{\,\mathrm{rest}}\,M_j^{(l)}(x),
\end{equation}
and update the gate as a small unitary step:
\begin{equation}
U_j^{(l)} \leftarrow \exp\!\big(i\,\eta\,K_j^{(l)}\big)\,U_j^{(l)}.
\end{equation}
This preserves unitarity while following the local gradient direction implied
by the commutator form. 
Appendix~\ref{AppendixGrad}.

\subsection{Resource Scaling and Efficiency}
We compare the sample complexity of adjoint-channel propagation against
full-state tomography. Let $m_l$ be the number of tunable gates at layer $l$,
$d_l$ the register dimension at layer $l$, and $n_{\mathrm{proj}}$ the number
of projective measurements per observable estimate. The number of copies
required per training round is
\begin{equation}
N_{\mathrm{copies}} = n_{\mathrm{proj}}\, \sum_{l=1}^{L+1} m_l\Big(4^{\,(\,m_{l-1}+1\,)} - 1\Big).
\end{equation}
For comparison, full-state tomography scaling is
\begin{equation}
N_{\mathrm{tom}} = n_{\mathrm{proj}} \sum_{l=1}^{L+1} m_l \cdot 2\Big[(d_l\,d_{l-1})^2 - 1\Big].
\end{equation}
These expressions highlight the advantage of adjoint-channel propagation over
tomography in terms of sample complexity for gradient estimation. A broader
discussion of efficiency trade-offs, including SWAP-trick fidelity estimation
and projection noise bounds, is provided in Appendix~\ref{AppendixC}.

\subsection{Hybrid Interface}
Hybrid feature vectors are extracted via expectation values
\[
z_r = \mathrm{Tr}(M_r X_L), \quad \mathbf{z}=(z_1,\dots,z_R),
\]
while gradients of features are given as
\[
\frac{\partial z_r}{\partial \theta}
= \mathrm{Tr}\!\left(M_r \frac{\partial X_L}{\partial \theta}\right).
\]
These features are passed to a RF or DNN classifier, with sensitivity
analysis aligned to benchmarking results in Section~\ref{numresult}.
The discussion of hybrid interface is further elaborated in Appendix~\ref{AppendixHybrid}

--------------------------------------------------------------------

\section{Efficiency Considerations}
\label{AppendixC}

We analyze efficiency trade-offs including SWAP-trick fidelity estimation,
projection noise bounds, sparsity effects, and simulator alignment costs. These
considerations highlight practical limits of quantum-inspired benchmarking and
guide deployment choices across Layer~1--3 of the Hierarchical Threat Detection
Framework.

\paragraph{SWAP-trick fidelity estimation.}
The SWAP trick provides an efficient method to estimate purity or overlaps
without full tomography. For a density operator $X$, we have
\[
\mathrm{Tr}(S\, X\otimes X)=\mathrm{Tr}(X^2),
\]
where $S$ is the SWAP operator acting on duplicate registers. This can be
implemented via controlled-SWAP circuits with bounded depth. The shot complexity
scales with the variance of the SWAP estimator, making it suitable for shallow
circuits and Layer~1 anomaly detection benchmarks.

\paragraph{Projection noise bounds.}
When estimating $p=\mathrm{Tr}(M X_L)$ via projective measurements with $N$
shots, the standard error satisfies
\[
\mathrm{SE}(p) \le \sqrt{\frac{p(1-p)}{N}}.
\]
This bound quantifies the statistical uncertainty inherent in finite sampling.
Adaptive measurement strategies or importance sampling can tighten the bound,
but the scaling remains $O(1/\sqrt{N})$. These limits are critical for Layer~2
intrusion confirmation, where reliable probability estimates are required.

\paragraph{Sparsity and topology impacts.}
Sparse entangler graphs (e.g., linear nearest-neighbor CZ) reduce circuit depth
and mitigate decoherence, but limit correlation expressivity. Dense entanglers
(e.g., all-to-all CNOT) increase representational power at the cost of higher
simulation and hardware resources. This trade-off directly affects Layer~3
multiattack classification benchmarks, where global correlations may be needed.

\paragraph{Simulator alignment costs.}
Different simulators (state-vector, tensor-network, hardware backends) incur
varying resource costs. State-vector simulators scale exponentially with qubit
count but provide exact results. Tensor-network simulators exploit sparsity and
low entanglement to reduce cost. Hardware backends introduce noise and require
error mitigation. Aligning simulator choice with benchmarking layer ensures
tractable experiments and reproducible results.

\begin{table}[h]
\centering
\caption{Efficiency considerations in quantum-inspired benchmarking}
\label{tab:efficiency}
\begin{tabular}{|p{1.1cm}|p{2cm}|p{5cm}|}
\hline
\textbf{Aspect} & \textbf{Formula / Definition} & \textbf{Implication and Layer Relevance} \\
\hline
SWAP-trick fidelity estimation &
$\mathrm{Tr}(S\, X\otimes X)=\mathrm{Tr}(X^2)$ &
Efficient purity/overlap estimation; bounded-depth circuits; suitable for Layer~1 anomaly detection. \\
\hline
Projection noise bounds &
$\mathrm{SE}(p) \le \sqrt{\tfrac{p(1-p)}{N}}$ &
Statistical error scales as $O(1/\sqrt{N})$; critical for Layer~2 intrusion confirmation requiring reliable probability estimates. \\
\hline
Sparsity and topology impacts &
--- &
Sparse entanglers reduce depth but limit correlations; dense entanglers increase expressivity but raise resource costs; trade-off most relevant for Layer~3 multiattack classification. \\
\hline
Simulator alignment costs &
--- &
State-vector simulators exact but exponential; tensor-network simulators efficient under low entanglement; hardware backends noisy but realistic; simulator choice must align with benchmarking layer. \\
\hline
\end{tabular}
\end{table}





--------------------------------------------------------------------
\section{Layer CP Maps and Adjoint Calculus}
\label{AppendixCP}

We derive the completely positive and trace-preserving (CPTP) layer maps via
Stinespring dilation, prove complete positivity and trace preservation, and
develop the adjoint-channel formalism used for backpropagation.

\subsection{Layer-to-Layer CPTP Map via Stinespring}

\paragraph{Definition.}
For layer $l$, the CPTP map is defined as
\[
\mathcal{E}_l(X_{l-1})
=
\mathrm{Tr}_{\mathrm{anc},\,l}\!\left(
U_l \,[X_{l-1}\otimes \ket{0}\bra{0}]\, U_l^\dagger
\right),
\]
where $U_l$ is the unitary acting on the system plus ancilla, and the ancilla
is initialized in $\ket{0}\bra{0}$.

\paragraph{Complete positivity (CP).}
For any extension $R \geq 0$ on
$\mathcal{H}_{l-1}\otimes \mathcal{H}_R$,
\[
(\mathcal{E}_l \otimes \mathrm{id}_R)(R)
=
\mathrm{Tr}_{\mathrm{anc},\,l}\!\left(
U_l [R \otimes \ket{0}\bra{0}] U_l^\dagger
\right) \geq 0,
\]
since unitary conjugation preserves positivity and partial trace is itself a
completely positive map.

\paragraph{Trace preservation (TP).}
Using cyclicity of trace and
$\mathrm{Tr}(\ket{0}\bra{0})=1$, we have
\[
\mathrm{Tr}\,\mathcal{E}_l(X_{l-1})
=
\mathrm{Tr}\,\mathrm{Tr}_{\mathrm{anc},\,l}\!\left(
U_l [X_{l-1}\otimes \ket{0}\bra{0}] U_l^\dagger
\right)
=
\mathrm{Tr}(X_{l-1}).
\]
Thus, $\mathcal{E}_l$ is CPTP and maps
$\mathcal{D}(\mathcal{H}_{l-1})$ to $\mathcal{D}(\mathcal{H}_l)$.

\subsection{Network Composition}

\paragraph{Forward composition.}
The full network map is
\[
\mathcal{E}_{1:L} = \mathcal{E}_L \circ \cdots \circ \mathcal{E}_1,
\qquad
X_L = \mathcal{E}_{1:L}(X_0).
\]

\paragraph{Prediction via measurement.}
Given observable $M$, the prediction is
\[
p = \mathrm{Tr}(M X_L).
\]

\subsection{Adjoint (Heisenberg) Maps}

For linear maps, the adjoint $\mathcal{E}_l^\dagger$ satisfies
\[
\mathrm{Tr}\big(A\,\mathcal{E}_l(B)\big)
=
\mathrm{Tr}\big(\mathcal{E}_l^\dagger(A)\,B\big)
\quad \forall A,B.
\]

\paragraph{Adjoint of the layer map.}
\[
\mathcal{E}_l^\dagger(A_l)
=
\mathrm{Tr}_{\mathrm{anc},\,l}\!\left(
U_l^\dagger [A_l\otimes I_{\mathrm{anc},\,l}]\, U_l
\right).
\]

\paragraph{Network adjoint composition.}
\[
\mathcal{E}_{1:L}^\dagger
=
\mathcal{E}_1^\dagger \circ \cdots \circ \mathcal{E}_{L-1}^\dagger \circ \mathcal{E}_L^\dagger.
\]

\paragraph{Backpropagation of observables.}
Starting from $A_L := M$, we propagate upstream:
\[
A_{l-1} = \mathcal{E}_l^\dagger(A_l), \qquad
p = \mathrm{Tr}(A_0 X_0).
\]
This adjoint-channel backpropagation yields gradients by propagating measurement
operators upstream through the network.





--------------------------------------------------------------------
\section{Gate-wise Gradients}
\label{AppendixGrad}

This section develops gate-wise gradient rules using generator calculus and
parameter-shift identities. Explicit formulas for derivatives of rotation and
entangling gates are provided, consistent with the adjoint-channel framework in
Appendix~\ref{AppendixCP}.

\subsection{Parameterized Gates}

Each parameterized gate is expressed as
\[
U_j^{(l)}(\theta) = e^{-i\theta H_j^{(l)}}, \qquad H_j^{(l)} = H_j^{(l)\,\dagger}.
\]

\subsection{Derivatives of a Gate}

The derivative of a gate with respect to its parameter is
\[
\frac{\partial U_j^{(l)}}{\partial \theta}
= -i\,H_j^{(l)} U_j^{(l)}, \qquad
\frac{\partial U_j^{(l)\,\dagger}}{\partial \theta}
= i\, U_j^{(l)\,\dagger} H_j^{(l)}.
\]

\subsection{Derivative of a Layer Map}

For a layer CPTP map $\mathcal{E}_l$, the derivative is
\beqn
\frac{\partial\,\mathcal{E}_l(X_{l-1})}{\partial \theta}
=
\mathrm{Tr}_{\mathrm{anc},\,l}\!\Big(
\frac{\partial U_l}{\partial \theta}
[X_{l-1}\otimes \ket{0}\bra{0}]\, U_l^\dagger
+\nonumber\\
U_l [X_{l-1}\otimes \ket{0}\bra{0}]\, \frac{\partial U_l^\dagger}{\partial \theta}
\Big).
\eeqn

The prediction derivative then follows as
\[
\frac{\partial p}{\partial \theta}
=
\mathrm{Tr}\!\left(
A_{l-1}\,
\frac{\partial\,\mathcal{E}_l(X_{l-1})}{\partial \theta}
\right),
\quad
A_{l-1}=\mathcal{E}_l^\dagger(A_l).
\]

\subsection{Chain Rule for Loss Functions}

For a loss $C=C(p,y)$ depending on prediction $p$ and target $y$, the derivative
is
\[
\frac{\partial C}{\partial \theta}
=
\frac{\partial C}{\partial p}\cdot \frac{\partial p}{\partial \theta}.
\]

\subsection{Parameter-Shift Rule}

For Pauli-generated rotations with eigenvalues $\pm \tfrac{1}{2}$, the
parameter-shift identity provides exact gradients:
\[
\frac{\partial p}{\partial \theta}
=
\frac{p(\theta+\tfrac{\pi}{2}) - p(\theta-\tfrac{\pi}{2})}{2}.
\]

This identity is compatible with the adjoint-channel calculus and is often
preferred for simulator or hardware evaluation.

\subsection{Multi-Parameter Layers}

For a layer unitary $U_l(\boldsymbol{\theta}_l)=\prod_{j=1}^{m_l} U_j^{(l)}(\theta_j)$,
the derivative with respect to $\theta_j$ is
\[
\frac{\partial U_l}{\partial \theta_j}
=
\left(\prod_{r<j} U_r^{(l)}\right)\,(-iH_j^{(l)}U_j^{(l)})\,
\left(\prod_{r>j} U_r^{(l)}\right).
\]

This expression allows gate-local gradient evaluation by combining forward
states and backward observables, reducing to commutator forms for local gates.





--------------------------------------------------------------------
\section{Hybrid DQNN--RF and DQNN--DNN Interfaces}
\label{AppendixHybrid}

This appendix describes the mapping of quantum outputs to classical feature
vectors via expectation values of fixed observables. The resulting features can
then be consumed by either a random forest (RF) classifier or a deep neural
network (DNN), providing hybrid quantum--classical pipelines for threat
detection. The RF interface emphasizes robustness and interpretability, while
the DNN interface emphasizes scalability and end-to-end differentiability.

\subsection{Feature Extraction and Gradients}
Given a set of observables $\{M_r\}_{r=1}^R$ acting on the final layer state
$X_L$, we define
\[
z_r = \mathrm{Tr}(M_r X_L), \qquad \mathbf{z}=(z_1,\dots,z_R).
\]
The vector $\mathbf{z}$ serves as the classical representation of the quantum
state. Gradients of features with respect to circuit parameters $\theta$ are
given by
\[
\frac{\partial z_r}{\partial \theta}
= \mathrm{Tr}\!\left(M_r \frac{\partial X_L}{\partial \theta}\right).
\]
These derivatives quantify the sensitivity of each feature to quantum parameters
and can be aggregated to identify which observables contribute most strongly to
classification performance.

\subsection{Integration with Random Forest}
The hybrid feature vector $\mathbf{z}$ is passed to a random forest classifier.
Random forests consume these features without requiring differentiability,
making them robust to noise and suitable for deployment in adversarial settings.
Sensitivity analysis of $\partial z_r/\partial \theta$ aligns with benchmarking
results in Section~\ref{numresult}, where feature importance is compared across
Layer~1 anomaly detection, Layer~2 intrusion confirmation, and Layer~3
multiattack classification.

\subsection{Integration with Deep Neural Networks}
Alternatively, the hybrid feature vector $\mathbf{z}$ can be passed to a deep
neural network (DNN). Unlike random forests, DNNs exploit differentiability and
can be trained jointly with the quantum circuit using gradient backpropagation.
This integration supports end-to-end learning and scalability, complementing the
robustness of the RF interface.

For a gate parameter $\theta$, the gradient of features can be expressed more
explicitly using adjoint recursion:
\[
\frac{\partial z_r}{\partial \theta}
= \mathrm{Tr}\!\left(A_0^{(r)} X_0\right), \qquad
A_0^{(r)} = \mathcal{E}_{1:L}^\dagger(M_r)'_{\theta},
\]
where the prime indicates dependency via the layer where $\theta$ acts,
computed by
\[
A_{l-1}^{(r)} = \mathcal{E}_l^\dagger(A_l^{(r)}).
\]
The magnitudes $\big|\tfrac{\partial z_r}{\partial \theta}\big|$ can be
aggregated across gates or layers to identify which features depend most
strongly on quantum parameters. This aggregation supports ablation studies and
aligns with the toggles and metrics reported in the benchmarking tables of
Section~\ref{numresult}.

\end{document}